\documentclass[11pt, reqno]{article}
\usepackage{epsfig}
\usepackage{amssymb}
\usepackage{amsmath}
\usepackage{mathrsfs}
\usepackage{cite}
\usepackage{hyperref}
\usepackage{multirow}

\def\be{\begin{equation}}
\def\ee{\end{equation}}
\def\ba{\begin{eqnarray}}
\def\ea{\end{eqnarray}}
\def\l{\langle}
\def\r{\rangle}

\def\nl{\nonumber\\}
\def\ZZ{\mathcal{Z}}

\makeindex
\begin{document}

\begin{titlepage}

\begin{center}

{\Large \bf Notes on the scattering amplitude -- Wilson loop duality \\[5pt]}

\vspace{0.3cm}

{\bf S. Caron-Huot$^a$}

\vspace{.1cm}

{\it $^{a}$ School of Natural Sciences, Institute for Advanced Study, Princeton, NJ 08540, USA}

\vspace{0.5cm}
\end{center}

\begin{abstract}
We consider the duality between the four-dimensional S-matrix of planar maximally supersymmetric Yang-Mills theory
and the expectation value of polygonal shaped Wilson loops in the same theory.
We extend the duality to amplitudes with arbitrary helicity states by introducing a suitable supersymmetric extension of the Wilson loop.
We show that this object is determined by a host of recursion relations, which are valid at tree level and at loop level for a certain ``loop integrand" defined
within the Lagrangian insertion procedure.
These recursion relations reproduce the BCFW ones obeyed by tree-level scattering amplitudes,
as well as their extension to loop integrands which appeared recently in the literature,
establishing the duality to all orders in perturbation theory.
Finally, we propose that a certain set of finite correlation functions can be used to compute all first derivatives of the logarithm
of MHV amplitudes.
\end{abstract}

\bigskip
\bigskip

\end{titlepage}

\tableofcontents
\newpage 

\section{Introduction}

\subsection{Dual conformal symmetry and Wilson loops}

Among all quantum field theories in four spacetime dimensions, maximally supersymmetric Yang-Mills (SYM) theory, in the 't Hooft limit of a large number of colors,
is a most remarkable one.  For one thing, it is the first and still only known example of an integrable theory in four spacetime dimensions.
This integrability has been exploited successfully, so far, in the context of computing the spectrum of anomalous dimensions of the theory \cite{minahan,BES}.
This theory is also singled out as being the prototypical example of the AdS/CFT duality \cite{maldacena},
which has produced many insights from strong coupling that have been instrumental in these developments.

Integrability structures arise in the study of scattering amplitudes.  By ``integrability'' we mean an infinite dimensional
symmetry algebra acts on amplitudes.  Most of these symmetries are hidden, non-local, symmetries that are not obvious
from the usual presentation of the theory.
The simplest example of such a symmetry can be seen from the expression of the one-loop correction to the four-point 1-loop amplitude.
Ignoring regularization for a moment, this can be written, for a color-ordered partial amplitude, as \cite{green}
\be
 A_4^{\textrm{1-loop}} = A_4^{\textrm{tree}} \times ig^2N_c 
 \int \frac{d^4q}{(2\pi)^4} \frac{(p_1+p_2)^2(p_2+p_3)^2}{q^2(q+p_1)^2(q+p_1+p_2)^2(q-p_4)^2}.
\ee
Besides being very compact, this expression exhibits a very remarkable symmetry.
This becomes manifest when one introduces the so-called dual coordinates $x_i$ which are defined implicitly through
\be
 p_i^\mu=x_i^\mu-x_{i{-}1}^\mu,
\ee
in terms of which the integral becomes simply
\be
  \int \frac{d^4x}{(2\pi)^4} \frac{(x_1-x_3)^2(x_2-x_4)^2}{(x-x_1)^2(x-x_2)^2(x-x_3)^2(x-x_4)^2}.
\ee
The remarkable symmetry is the inversion $x_i\to x_i/x_i^2$, which sends $(x_i-x_j)^2\to (x_i-x_j)^2/x_i^2x_j^2$:  Equivalently,
the integrand is conformally invariant.  This ``dual conformal symmetry" acts in momentum space and is therefore quite
distinct from the familiar conformal symmetry of $\mathcal{N}=4$, whose arena is the original spacetime.

Further evidence for this hidden symmetry accumulated in the course of higher-loop computations, where it was repeatedly found that all integrals appearing in loop
corrections shared this property \cite{4ptDCI,magic}.  Of course, scattering amplitudes are infrared divergent,
and regularization, for instance using dimensional regularization, will break the symmetry.  So in the end most of the symmetry is broken at loop level.
Nevertheless, the breaking of the (bosonic) symmetry occurs in a very restricted and controlled fashion
which leads to powerful constraints on the answer.  Specifically, the breaking is entirely captured by the so-called Bern-Dixon-Smirnov (BDS)
Ansatz \cite{BDS}. Upon dividing by this Ansatz, any (finite) remainder is exactly conformal invariant \cite{4pt5pt}.
In particular, the BDS Ansatz provides the full all-loop answer for 4- and 5-points amplitudes,
because nontrivial conformal invariants do not exist in these cases.

In parallel development Alday and Maldacena \cite{aldaymaldacena1} studied scattering amplitudes at strong coupling using AdS/CFT.
They confirmed the BDS Ansatz at 4- and 5-points but observed deviations for sufficiently large number of particles \cite{aldaymaldacena2};
deviations at 6 points were also deduced from investigation of Regge limits \cite{LipatovRegge}. These violations are described by conformal invariant
remainders, consistent with the symmetry analysis just mentioned.

The strong coupling analysis led to a very geometrical understanding of the dual symmetry:
a certain T-duality \cite{maldacenaberkovits} takes planar $\mathcal{N}=4$ to itself, loosely speaking interchanging the dual coordinates and space-time coordinates.
It converts a scattering process to the expectation value of a polygonal Wilson loop with vertices at the points $x_i$,
\be
 M_n(p_1,\ldots,p_n) \Leftrightarrow W_n(x_1,\ldots,x_n).
\ee
Successive segments of the polygonal contour have length $p_i^\mu$;
we refer the reader to \cite{aldayroiban} for an excellent review.
The ``hidden'' symmetries of the scattering amplitudes have now become ordinary conformal transformations of this Wilson loop.

It makes sense to compute the same Wilson loop at weak coupling.  The result was found, by explicit computations, to agree with scattering amplitudes,
but with a certain very  specific set of helicity configurations, the so-called MHV amplitudes \cite{WL1,brandhuberloop}.
This Wilson loop/MHV amplitude duality has been confirmed up to two loops for the six-point function \cite{hexagon1,hexagon2,hexagon3}.

In parallel development, it was understood how dual (super)conformal generators act on the polarization data
that defines a scattering amplitude, allowing to define the symmetry away from the MHV limit.  Tree amplitudes in
that case have been shown to be invariant under a full dual superconformal \cite{sokatchevDCI,brandhubernote,Hodges,masonDCI,ahgrassmannian},
while loops carry only a (mildly broken) dual conformal invariance \cite{sokatchevDCI,sokatchevsDCI}.

A certain small piece had thus been missing for some time in order to bring this story to its full completion: a geometrical understanding of the symmetry
for non-MHV amplitudes, similar to that afforded by the Wilson loop in the MHV case.
One of the aims of this paper is to provide a super-Wilson loop that fulfills just that purpose.%
\footnote{While this manuscript was being prepared, another Wilson loop, very likely to be equivalent, has been
 derived from a very different perspective by Mason and Skinner \cite{masonskinner}.}

\subsection{Loop integrands}

A significant extension of the scope of the duality was proposed recently in \cite{EKS2}.
There it was proposed that the duality holds not only for final answers, but
already for suitably defined integrands.  The relevant ``integrands'' have been defined in two recent papers, in two very different contexts.

In the context of Wilson loop computations, Eden, Korchemsky and Sokatchev \cite{EKS1} introduced a ``loop integrand''
by using the so-called method of Lagrangian insertion. In this method, loops are computed by taking repeated derivatives with respect to the coupling constant,
which in SYM brings down powers of the Lagrangian density.
At $\ell$ loops these are to be integrated over $\ell$ copies of space-time,
the integrand being given as a tree-level correlation function of the Wilson loop with $\ell$ Lagrangian insertions.
This way the computation of a Wilson loop is turned into a set of $d^4x$ integrals.
By adding suitable total derivatives to the action density, a ``chiral'' Lagrangian insertion involving
$F_{\alpha\beta}F^{\alpha\beta}$ instead of $F_{\mu\nu}F^{\mu\nu}$ can be used, which was found in \cite{EKS1} to be greatly convenient.

Independently, Arkani-Hamed and collaborators  \cite{abcct} including the present author introduced a loop integrand in the context of scattering amplitude computations.
At the simplest level, this  ``integrand'' is just the familiar momentum-space one, obtained after summing over all Feynman graphs and doing numerator algebra.
Normally, such an integrand is not quite well-defined, because loop momenta are only dummy integration variables: it is unclear
how to add different Feynman graphs under the same integration sign. One could always shift one diagram relative to the others.
This ambiguity turns out to be absent in the planar limit,
where region momenta $x_i$ can be used to canonically add diagrams.  This way a gauge-invariant integrand is obtained.
The work \cite{abcct} gave an efficient recursive procedure for computing it, starting from manifestly on-shell and physical tree amplitudes.

The two integrands are known empirically to produce the same integrals,
and it is therefore very natural to ask whether they are actually the same.
Both integrals are chiral (non parity-invariant): in the first case due to a technical choice,
in the second case due to one considering MHV amplitudes as opposed to its parity conjugate $\overline{\textrm{MHV}}$.
Thus they do not obviously disagree.  Remarkably, explicit comparison in \cite{EKS2} has revealed that they precisely agree!
In this paper we aim to provide an explanation for this fact.  The explanation will, in fact, lead directly to a Wilson loop with polarizations,
and to a general proof of the duality.

\subsection{Super-Wilson loop and recursion relations}

Our starting point is the singularities of the integrand defined by Lagrangian insertions in \cite{EKS1}.
This is a rational function of the insertion point $y$, with simple poles of the form $1/(y{-}x_i)^2$ where $y$ becomes
null-separated from a corner of the Wilson loop.  These singularities can be understood at the operator level: a singular
propagator knocks out one factor of $F_{\alpha\beta}$ from a Lagrangian insertion,
\be
 F^2 \sim \frac{1}{(y-x_i)^2} \lambda_{\alpha}\lambda_{\beta} F^{\alpha\beta}(y) + \mbox{non-singular}.
\ee
Here $\lambda$ is some spinor associated with the separation $(y-x_i)$ that is becoming null.
The resulting correlation function can be made gauge-invariant in a natural way, by adding two segments to the Wilson loop joining $x_i$ to $y$ and back.
Thus the residue at the pole is a Wilson loop with two more edges and $F_{\alpha\beta}$ inserted at a corner:
\be
\raisebox{-1.5cm}{\includegraphics[scale=2.5]{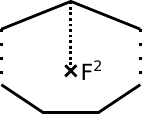}} \hspace{0.5cm} \to \hspace{0.5cm}
 \raisebox{-1.5cm}{\includegraphics[scale=2.5]{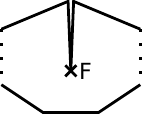}} \nonumber.
\ee
In the scattering amplitude context, exactly the same residue is interpreted as the forward limit of a NMHV tree amplitude
with two particles added \cite{loopandtrees,abcct}.
Thus Wilson loops, with suitable insertions, know about the forward limit of NMHV tree amplitudes!

As we will see, the duality is not limited to forward limits: with suitable insertions along the contour,
general NMHV tree amplitudes are also reproduced.
Furthermore, there is no reason to stop at NMHV level.  We will define a super-Wilson
loop that knows about all N${}^k$MHV amplitudes, and agrees with all MHV-stripped super-amplitude $M_n$ where
\be
 A_n\equiv \frac{g^{n-2} (2\pi)^4\delta^4(\sum p)\delta^8(\sum q)}{\l 12\r\l23\r\cdots\l n1\r} \times M_n \label{MHV}.
\ee

This new sweep of super-Wilson loops depends on the same
momentum and polarization data as scattering amplitudes do.   Importantly, within this enlarged sweep of objects, recursion relations become possible
that relate objects with $\ell$ Larangian insertions to other objects with fewer Lagrangian insertions but more edges.
These recursion relations will be seen to be identical to those found in \cite{abcct}, establishing that the two sweeps of objects are the same,
to all loops, for arbitrary polarizations, at the integrand level.

This paper is organized as follows. In section \ref{sec:momtwist} we introduce the coordinates that prove to be particularly well-suited
for our description of scattering amplitudes, the so-called momentum twistor variables.  In section \ref{sec:NMHV} we begin the construction
of the super-Wilson loop through a process of ``reverse-engineering the answer.'' We complete the construction in section \ref{sec:wloop} by requiring
that the object be supersymmetric, in a certain specific sense.  Then in \ref{sec:example} we give a simple example  confirming that the super-Wilson loop
is indeed able to reproduce 1-loop NMHV amplitudes.  The remainder of the paper is devoted to establishing the recursion relations,
at tree-level \ref{sec:tree} and for loop integrands \ref{sec:loop}.  The paper ends with some conclusions, together with Appendices
that give more details about the Wilson loop operator and various computations.  In a final Appendix we outline, based on physical arguments,
how divergences are expected to factorize within the Lagrangian insertion procedure.

\section{Lightning introduction to momentum twistors}
\label{sec:momtwist}

There are many useful ways to describe tree amplitudes in $\mathcal{N}=4$ SYM.
Since we are interested in the dual Wilson loop picture, we should employ those
variables that make the symmetries of the Wilson loop most manifest.  Incidentally, these variables are also
the ones which produce the most compact analytic expressions:
These are the momentum (super-)twistor variables introduced by Hodges \cite{Hodges}. 

Momentum twistors can be used in any planar theory with massless particles.
Their main achievement is to explicitly solve both the momentum conservation and mass-shell condition constraints, allowing
for a constrain-free description of the external data.  The constrains can be solved in two successive steps.
The first step was already implicit in the introduction: one can trivialize momentum conservation in a planar theory by introducing variables $x_i$
such that
\be
 p_i^\mu= x_i^\mu - x_{i{-}1}^\mu.  \label{xspace}
\ee
Planarity provides the required ordering of the particle labels.
In the scattering amplitude picture the $x_i$ are called region momenta, while in the dual theory they are the positions of the corners
of the Wilson polygon.

To solve the mass-shell constraints $(x_i - x_{i{-}1})^2=0$ we proceed to assign a ``momentum twistor'' to each particle.
This is a four-component object
\be
 Z_i^\alpha \equiv \left(\begin{array}{c} \lambda_i^\alpha \\ \mu_i^{\dot\alpha}\end{array}\right)
= \left(\begin{array}{c} \lambda_i^\alpha \\ x_{i\alpha}{}^{\dot\alpha}\lambda_i^\alpha \end{array}\right).
\ee
where the spinor $\lambda_i^\alpha$ is the one entering the standard decomposition $p_i^{\alpha\dot\alpha}=\lambda_i^\alpha\tilde\lambda_i^{\dot\alpha}$,
for $p_i^2=0$.  $\lambda_i^\alpha$ is defined only projectively and so is $Z_i^\alpha$: $Z_i^\alpha$ and $t_i Z_i^\alpha$ are to be identified, for
any nonzero complex number $t_i$.

The fact that this does solve the on-shell conditions will be explained shortly.  As a first check one can count the degrees of freedom.
We have $4n$ momentum twistor components, subject to $n$ projective invariances and $4$ translation invariances for the $x_i$, leaving $(3n-4)$ degrees of freedom.
The original scattering data, on the other hand, has $4n$ variables subject to $n+4$ constraints, giving also $(3n-4)$ degrees of freedom.
Thus all constraints have become symmetries.

The inverse change of variable can be constructed explicitly \cite{Hodges}:
\be
 x_i^{\alpha\dot\alpha} = \frac{\lambda_i^\alpha \mu_{i{+}1}^{\dot\alpha} - \lambda_{i{+}1}^\alpha \mu_{i}^{\dot\alpha}}{\l i\,i{+}1\r}.
\ee
The symbol $\l ab\r$ is the standard antisymmetric 2-bracket, defined using the upper two components $\lambda_{a,b}^\alpha$ of $Z_a$ and $Z_b$
(and more precisely defined in Appendix).
Another useful equation is
\be
 \tilde\lambda_i^{\dot\alpha} = \frac{
 \l i{-}1\,i\r \mu_{i{+}1}^{\dot\alpha}
 +\l i\,i{+}1\r \mu_{i{-}1}^{\dot\alpha}
 +\l i{+}1\,i{-}1\r \mu_{i}^{\dot\alpha}}{\l i{-}1\,i\r\l i\,i{+}1\r},
\ee
while the Lorentz-invariant distance is simply
\be
 (x_i-x_k)^2= \l i\,i{+}1\r \l k\,k{+}1\r \times  \l i\,i{+}1\,k\,k{+}1\r
\ee
where $\l abcd\r$ is the totally antisymmetric contraction $\epsilon_{i_1i_2i_3i_4} Z_a^{i_1}Z_b^{i_2}Z_c^{i_3}Z_d^{i_4}$.

Geometrically, the point $x_i$ in Minkowski space is identified with the line $(i\,i{+}1)$ in momentum twistor space, that passes through
the points $Z_i$ and $Z_{i{+}1}$.  The condition that $x_i$ and $x_j$ be null separated is that $\l i\,i{+}1\,j\,j{+}1\r=0$.
Geometrically, this happens if and only if the lines $(i\,i{+}1)$ and $(j\,j{+}1)$ intersect.

We can now see why the mass-shell conditions are solved: the lines $(i\,i{-}1)$ and $(i\,i{+}1)$ automatically intersect, the intersection
being just the point $Z_i$.  Thus $p_i^2\equiv (x_i-x_{i{-}1})^2=0$ for any value of the momentum twistors.

Momentum twistors make conformal properties manifest, even though conformal invariance is not a prerequisite for their use.
The conformal group in four dimensions is SO(4,2) which is realized linearly on the momentum twistors.  More precisely, momentum twistors transform
in the spinor representation of (the double-cover) of SO(4,2).
Recall that we are describing the dual conformal transformations, which are those acting simply in the $x_i$ coordinates of the Wilson loop
and not to be confused with the ordinary conformal transformations (acting on the original spacetime where the scattering process occurs).

Because the antisymmetric tensor $\epsilon_{ABCD}$ is invariant, any expression
made out of 4-brackets is conformally invariant, while any 2-bracket breaks conformal invariance.

To describe $\mathcal{N}=4$ polarization states one introduces the expansion for the on-shell states  \cite{nair}
\be
   |i\rangle = |g^+_i\rangle + \tilde\eta_i^A |\psi_{iA}\rangle + \frac12 \tilde\eta_i^A \tilde\eta_i^B |\phi_{iAB}\rangle
      + \frac1{3!} \epsilon_{ABCD} \tilde\eta_i^A\tilde\eta_i^B\tilde\eta_i^C |\tilde\psi_i^D\rangle + \frac{1}{4!} \tilde\eta_i^4 |g^-_i\rangle .
\ee
(Notice our nonstandard use of tildes in this expansion.  Untilded $\eta$-variables will be defined shortly.)
Capitalized indices range from $1$ to $4$ and carry the SU(4)${}_R$ R-symmetry charge,
yielding two gluon helicities, four chiral fermions, four antichiral fermions and 6 scalars.

Similarly to the $x_i$ coordinates, one can introduce Grassman anticommuting
coordinates $\theta_i$ such that $\lambda_i^\alpha \tilde\eta_i^A = \theta_{i}^{\alpha A}-\theta_{i{-}1}^A$.
Momentum super-twistors are then defined as
\be
 \mathcal{Z}^\alpha = \left(\begin{array}{c} \lambda_i^\alpha \\ \mu_i^{\dot\alpha}\\ \eta_i^A\end{array}\right)
= \left(\begin{array}{c} \lambda_i^\alpha \\ x_{i\alpha}{}^{\dot\alpha}\lambda_i^\alpha \\ \theta_{i\alpha}^A \lambda_i^\alpha \end{array}\right).
\ee
The inverse map is
\be  \label{etachange}
 \tilde\eta_i^{A} = \frac{
 \l i{-}1\,i\r \eta_{i{+}1}^{A}
 +\l i\,i{+}1\r \eta_{i{-}1}^{A}
 + \l i{+}1\,i{-}1\r \eta_{i}^{A}}{\l i{-}1\,i\r\l i\,i{+}1\r}.
\ee

Similar to the $x_i$, the $\theta_i$ trivialize super-momentum conservation $\sum_i \lambda_i \tilde\eta_i$, which is part of the supersymmetry algebra.
A consequence is that $\theta$ and $\eta$ only make sense on the support of suitable super-momentum conserving delta functions.
That is why, whenever one works with super-momentum twistors,
one has to work with the MHV-stripped stripped amplitudes $M_n$ introduced above.
Dual conformal invariance of the tree-level scattering amplitudes in $\mathcal{N}=4$ amounts to $M_n$ being expressible,
using $Z_i$ and $\eta_i$ variables, using only four-brackets.

\section{NMHV tree amplitudes from Wilson loops}
\label{sec:NMHV}

The (super) momentum twistors make tree amplitudes
extremely well localized in the coordinates $x_i$; this is the coordinate space of Wilson loop and so this is particularly relevant for us.
For instance, consider the NMHV super-amplitude.  It admits the very compact form \cite{sokatchevDCI,Hodges}.
\be
 M_n^{\textrm{NMHV}} = \sum_{1< i< j < n-1} [1\,i\,i{+}1\,j\,j{+}1] \label{NMHV}
\ee
where the square bracket is the ubiquitous supersymmetry invariant (Yangian invariant)
\be
 [a\,b\,c\,d\,e] = \frac{\delta^{0|4}(\eta_{a} \langle b \, c \, d \, e \rangle + {\rm cyclic})}
 {\langle a \, b \, c \, d\rangle \langle b \, c \, d \, e \rangle \l c\,d\,e\,a \r\l d\,e\,a\,b\r \langle e\, a \, b \, c \rangle}.
\ee

With simple choices of $\tilde\eta_i$  the same expression gives NMHV tree amplitudes in pure Yang-Mills theory.
The amplitudes with such values of $\tilde\eta$ are nontrivial linear combinations of those with prescribed $\eta$ values, and the conversion
between $\eta$- and $\tilde\eta$- amplitudes requires full $\mathcal{N}=4$ amplitudes.  In that sense the duality under consideration
requires the full $\mathcal{N}=4$ supersymmetry:
it is those amplitudes with prescribed $\eta$ that will appear on the Wilson loop side.

To see how $\eta$-components localize along the Wilson loop consider for instance
the component proportional to $\eta_1\eta_2 \eta_k\eta_{k{+}1}$ in the scattering amplitude.  It comes from only one term\footnote{
 To avoid cluttering the notation, here we omit the SU(4)${}_{\mbox{R}}$ indices on the variables $\eta_i^A$.
 Whenever we write a product of four $\eta$'s, it is understood that all $\eta$'s have different $R$-charge.
 Here, for instance, we could be extracting the component $\eta_1^1\eta_2^2\eta_k^3\eta_{k{+}1}^4$.},
\be
 [1\,2\,3\,k\,k{+}1] \simeq \eta_1\eta_2\eta_k\eta_{k{+}1} \times \frac{1}{\langle 1\,2\,k\,k{+}1\rangle}. 
\ee
Manifestly, this is completely localized to the corners $x_1$ and $x_k$ of the Wilson loop:
by inserting scalar field operators at corners of the Wilson loop
\be
 \frac{\phi_{AB}(x_i)}{\l i\,i{+}1\r} \eta_i^A\eta_{i{+}1}^B
\ee
and connecting two corners with a scalar propagator this will be reproduced
\be
  \frac{\eta_1\eta_2\eta_k\eta_{k{+}1}}{\l12\r\l k\,k{+1}\r(x_1-x_k)^2}
=
 \eta_1\eta_2\eta_k\eta_{k{+}1} \times \frac{1}{\l1\,2\,k\,k{+}1\r}.
\ee
\be
 \raisebox{-1.5cm}{\includegraphics[scale=1.0]{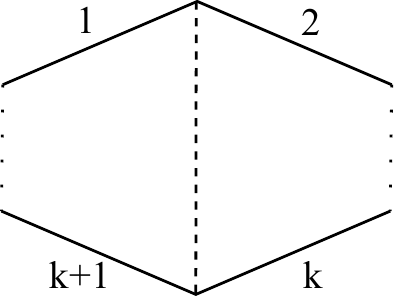}} \nonumber.
\ee

This completely fixes the dependence of a putative super-Wilson loop on the product $\eta_i\eta_{i{+}1}$, if the tree NMHV amplitude
is to be reproduced.  Any other dependence on $\eta_i\eta_{i{+}1}$ could not reproduce this simple result.

Other simple lessons gathered from the NMHV formula are that no dependence can occur on products such as $\eta_i \eta_{i{+}2}$
(because this would generate terms not in (\ref{NMHV})),
and that any operator inserted on the edge $p_i$ can depend only on the single momentum twistor $\eta_i$.

To find out what these edge insertions are for scalar field components, consider the component $\eta_2^2\eta_k\eta_{k{+}1}$
of the super-amplitude
\be
 [1\,2\,3\,k\,k{+}1] \sim \eta_2^2\eta_k\eta_{k{+}1} \times \frac{\l 1\,3\,k\,k{+}1\r}{\l 1\,2\,k\,k{+}1\r\l2\,3\, k\,k{+}1\r}.  \label{inducedphi}
\ee
According to the above analysis we must reproduce this using a propagator linking the corner $(k\,k{+}1)$ to
the edge $p_2$, or its endpoints.
A natural Ansatz for the edge operator on segment 2, which will be motivated shortly, is
\be
 \frac14 \int_0^1 dt  \frac{\tilde\lambda_2 v D \phi_{AB}(x_1+tp_2)}{\l 2v\r}  \eta_2^A\eta_2^B.
\ee
The derivative is required to match with the dimensionality of the corner term and
its form will be explained in a moment.  For the moment
this Ansatz contains an undetermined spinor $v^\alpha$.   To determine it we compute the partial amplitude
$\eta_2^2\eta_k\eta_{k{+}1}$ it produces:
\be
  \frac12\int_0^1 dt \frac{\tilde\lambda_2 v \partial}{\l 2v\r \l k\,k{+}1\r} \frac{1}{(x_1+tp_2-x_k)^2}
 = \frac{\tilde\lambda_2 v (x_k-x_1)}{\l v2\r \l k\,\,k{+}1\r (x_1-x_k)^2(x_2-x_k)^2}.
\ee

This conformal properties of this expression are not obvious.  To make such properties manifest
we rewrite the right-hand side using momentum twistor variables:
\be
 \frac{ \l v\,k\,k{+}1\,\,[1\r \l 2\,3]\r}{\l 1\,2\,k\,k{+}1\r\l 2\,3\,k\,k{+}1\r \l v2\r}
\ee
where the square bracket means cyclic sum.
It now becomes apparent that the choice $v^\alpha=\lambda_1^\alpha$ produces almost the right answer:
\be
 \frac{ \l 1\,k\,k{+}1\,\,[1\r \l 2\,3]\r}{\l 1\,2\,k\,k{+}1\r\l 2\,3\,k\,k{+}1\r\l 12\r}
  = \frac{\l 1\,3\,k\,k{+}1\r}{\l 1\,2\,k\,k{+}1\r\l2\,3\,k\,k{+}1\r}
  + \frac{\l 31\r}{\l 2\,3\,k\,k{+}1\r \l12\r},
\ee
where we have used the Schouten identity.
The remainder is localized on the corner $(23)$, and can be dealt with by a suitable corner term.

Some comments are in order.  First, the integral has produced a rational function of the external data (e.g., there is no logarithm).
This is surprising because integration often produces logarithms.  Here this can be traced to the chiral derivative acting on $\phi$.
In fact,  a rational answer will only be obtained for derivatives of the form $\tilde\lambda_{2\dot\alpha}D^{\alpha\dot\alpha}$
or $\lambda_{2\alpha} D^{\alpha\dot\alpha}$.  Investigation of the latter case reveals that it is incapable of producing the correct NMHV answer,
leaving the above Ansatz as the only possibility.

Therefore what we have found so far is that
the coupling of the Wilson loop to scalar fields can be uniquely determined by ``reverse-engineering'' the NMHV formula:
\be
 \sum_i \frac14 \int_0^1 dt \frac{\tilde\lambda_i \lambda_{i{-}1} D\phi_{AB}(x_{i{-}1}+tp_i)}{\l i\,i{-}1\r} \eta_i^A\eta_i^B
 + \phi_{AB}(x_i) \frac{\eta_i^A}{\l i\,i{+}1\r} \left( \eta_{i{+}1}^B - \frac12 \frac{\l i{-}1 i{+}1\r \eta_i}{\l i{-}1\,\,i\r}\right). 
\ee


It is clear that by considering more and more polarizations we could continue with this ``reverse-engineering'' process and determine
couplings to fermions and gauge boson.  The tree-level NMHV analysis, however, will always be determined by free-theory two-point functions
and is incapable of constraining non-linear couplings.
To fully determine the super Wison-loop this way,
we would have to consider N${}^2$MHV tree amplitudes, or NMHV loop amplitudes, and so on.
Fortunately there is a better way to proceed.

\section{The full Wilson loop from supersymmetry constraints}
\label{sec:wloop}

Tree-level scattering amplitudes are known to be invariant under dual super-conformal transformations.
They are in fact invariant under a whole Yangian algebra, but
here we should focus only on those symmetries which the Wilson loop formulation is expected to make manifest.
These contain the Poincar\'e supersymmetries $q,\tilde q$ of the Wilson loop which are expected to match certain elements of the SU(4$|$4) supersymmetries
acting on momentum twistors.
Specifically, the expected correspondence is obtained by matching the quantum numbers:
\ba
 q_A^\alpha &\Leftrightarrow& \sum_i \lambda_i^\alpha \frac{\partial}{\partial \eta_i^A} = \sum_i \frac{\partial}{\theta^A_{\alpha i}}, \\
 \tilde q^{A\dot\alpha} &\Leftrightarrow& \sum_i \eta_i^A \frac{\partial}{\partial \mu_i^{\dot\alpha}} = \sum_i \theta_i^{A\alpha} \frac{\partial}{\partial x_i^{\alpha\dot\alpha}}.
\ea
(Notice that the R-charges carried by elementary fields are opposite in the T-dual theory.)
The theory is also invariant under super-conformal generators
$s^{A\alpha}$ and $\tilde s_A^{\dot\alpha}$, which will be discussed below.

It is useful to note that $q_A$ does not actually act on the original space-time scattering data: from that viewpoint $q_A$ 
is simply a redundancy in the $\theta$ variables, much like translation invariance of the $x_i$.
Thus the super-Wilson loop has better be invariant under these transformations.
However, contrary to translation symmetry, the bosonic part $\oint dx^\mu A_\mu$ is not trivially invariant under $q_A$:
this invariance is a very nontrivial constraint.

To proceed we write the general couplings in the form
\be
 W_n\equiv \frac{1}{N_c} \mbox{Tr} \left[ {\mathcal P} 
 e^{\int_0^1 dt \mathcal{E}_1(t)} 
 \,\,\mathcal{V}_{12} 
 \,\,{\mathcal P} e^{\int_0^1 dt \mathcal{E}_2(t)} 
 \,\,\mathcal{V}_{23}  \cdots 
  \,\,{\mathcal P} e^{\int_0^1 dt \mathcal{E}_n(t)} \mathcal{V}_{n1}\right] \label{wloop}
\ee
where the $\mathcal{E}_i(t)\equiv \mathcal{E}_i(x_{i{-}1} + t p_i)$ are edge operators and the $\mathcal{V}_{i\,i{+}1}$ are vertex operators.
In line with the above analysis, we will impose that $\mathcal{E}_i$ depends only on $\eta_i$,
and that $\mathcal{V}_{i\,i{+}1}$ depend only on $\eta_i$ and $\eta_{i{+}1}$.  Ultimately this is justified by the fact that a solution will be found
with these constraints.

\def\EE{\mathcal{E}}

The edge operator begins with the gauge connection $\mathcal{E}_i = -p_i{\cdot} A$.  The variation of this term is nonzero, as said above,
$q_A^\alpha p_i{\cdot}A =\frac12 \lambda_i^\alpha \tilde\lambda_{i\dot\beta} \tilde\psi_A^{\dot\beta}$.  This forces us
to add a fermion term such that $(q_A^\alpha + c_0 \sum \lambda^\alpha \frac{\partial}{\partial \eta^A})W=0$:
\be
 \EE_i = -p_i{\cdot}A + \frac1{2c_0} \tilde\lambda_{i\dot\beta} \tilde\psi_A^{\dot\beta} \eta_i^A + \ldots
\ee
This is unique if we assume that $\EE_i$ depends only on $\eta_i$.

\subsection{The Wilson loop in component form}

The procedure is systematic and by repeating it one obtains terms of higher and higher order in $\eta$. With some algebra we find
\ba \hspace{-0.5cm}
 2\mathcal{E}_i &=& -\tilde\lambda_i \lambda_i A
 + \frac1{c_0} \tilde\lambda_i \tilde\psi_A \eta_i^A
 + \frac1{2c_0^2} \frac{\tilde \lambda_i \lambda_{i{-}1} D \phi_{AB}}{\l i\,i{-}1\r} \eta_i^A\eta_i^B \nl
 &&
 - \frac{1}{3!c_0^3} \epsilon_{ABCD}\frac{\tilde\lambda_i \lambda_{i{-}1}\lambda_{i{-}1} D\psi^A}{\l i\,i{-}1\r^2} \eta_i^B\eta_i^C \eta_i^D
 - \frac{1}{4!c_0^4} \epsilon_{ABCD}\frac{\tilde\lambda_i \lambda_{i{-}1}\lambda_{i{-}1}\lambda_{i{-}1} DF}{\l i\,i{-}1\r^3}\eta_i^A\eta_i^B\eta_i^C\eta_i^D,
  \label{edge}
\ea
together with the vertex terms
\be
 \mathcal{V}_{i\,i{+}1} = 1+ \frac{\phi_{AB}}{c_0^2\l i{+}1\,i\r} \left(\eta_{i{+}1}^A\eta_i^B - \frac12\frac{\l i{-}1\,i{+}1\r}{\l i{-}1\,i\r} \eta_i^A\eta_i^B\right) + \ldots
\ee
The somewhat lengthy expression for the vertex term is reported in Appendix.

The overall factor of $2$ in the edge term is just from $dx^\mu A_\mu \to \frac{dx_{\alpha\dot\alpha}}{2} A^{\alpha\dot\alpha}$:
for simplicity all terms are written as spinor products.  In each term there is a unique way to contract the indices, taken conventionally
to be with a lower spinor index on the left and upper index on the right.

The constant $c_0$ has appeared because the two supersymmetries act on different
superspaces: it is not \emph{a-priori} obvious how to normalize one compared to the other.
The correct value can be obtained by matching with the NMHV tree amplitude.  With careful normalization (which was omitted in the previous section)
we find
\be
 c_0=\left(\frac{g^2N_c}{4\pi^2}\right)^{1/4}. \label{defc0}
\ee

Our construction and verification that the Wilson loop is invariant under the claimed $q_A^\alpha$ supersymmetry has proceeded by brute force application
of the supersymmetry generators given in Appendix.  More precisely, the edges transform by total derivatives, which are cancelled by vertex variations.
The precise form of the total derivative is given in Appendix.
It is also important to use the equations of motion.

\subsection{Symmetries}

What about other symmetries?  Consider the generator $\tilde s_A$.  It is easy to compute the lowest term in the edge variation:
\be
(\tilde s_A^{\dot\alpha} + c_0 \mu_i^{\dot\alpha}\frac{\partial}{\partial \eta_i^A})( -\tilde\lambda_i \lambda_i A
 + \frac1{c_0} \tilde\lambda_i \tilde\psi_A \eta_i^A + \ldots)
  = -(\lambda_{i\beta} x^{\dot\alpha\beta}) \tilde\lambda_i \tilde\psi_A - \mu^{\dot\alpha} \tilde \lambda_i \tilde\psi_A + \ldots = 0+ \mathcal{O}(\eta).
\ee
Thus the lowest term of the variation vanishes; here we have used the definition of $\mu^{\dot\alpha}$.   Since the higher-order terms in $\eta$
are obtained by repeated action of $q_A$, and since $q_A$ and $\tilde s_A$ commute, this can be used
to show that the full operator is invariant.
More precisely, $\tilde s_A W$ is annihilated by $q_A$, vanishes in lowest degree,
and depends on edge $i$ only through $\eta_i$.  But there is no such thing, so $\tilde s_A W=0$ to all orders in $\eta$.

The same argument shows that the operator is (formally) conformal invariant.  In fact there is nothing to compute, since we already
know that the Wilson loop is conformal invariant  to order $\eta^0$.
So $K^{\alpha\dot\alpha} W$ vanishes in lowest component, is annihilated by $q$ (using that $[q_A,K]\propto \tilde s_A$ and that $\tilde s_A$ is already known
to be a symmetry), and depends on edge $i$ only through $\eta_i$: this implies $KW=0$.  We say ``formally'' because, contrary to $q_A$ and $\tilde s_A$,
$K^{\alpha\dot\alpha}$ contains bosonic derivative which make it more singular at short distances.  Indeed it is well-known that $K^{\alpha\dot\alpha}$ becomes anomalous at loop level
\cite{4pt5pt}; $q_A$ and $\tilde s_A$ do not become anomalous, however.

Finally, one might ask whether the operator is well defined, given that it contains null-separated fields.
It appears that built-in cancellation mechanisms prevent ill-defined contractions of null-separated fields from occurring in it.
For instance, the scalar field component $\phi(y)$ proportional to $\eta_2\eta_2$ that is induced from edge 2 and vertex 2, as seen from (\ref{inducedphi}), vanishes when
$\phi$ becomes null-separated from the endpoints of edge 2.  Thus no singular contraction involving this component of the
is possible.  In other cases supersymmetric cancellations can be seen.
Indeed a formal argument based on supersymmetry can be given to argue the absence of problematic contractions:
by using 12 of the supercharges $Z^\alpha \frac{\partial}{\partial \eta^A}$ to set
three adjacent $\eta^A$ to zero, any potential power-divergent contraction between the two vertices will be prevented.  Now of course one could do this anywhere around the loop,
suggesting no such contraction occurs anywhere.
Another line of argument is that the operator is essentially generated by repeated action
of $q_A$ on the bosonic Wilson loop:  since one never worries about the power-divergent contractions of $A^\mu$ (only about logarithimc divergent contributions at loop level),
this leaves little reason to worry about such contractions for the super-extension of the operator.  For these reasons we will discuss such issues any longer.

In summary, we have defined a super-Wilson loop operator that is manifestly invariant under $q_A^\alpha$ and $\tilde s_A^{\dot\alpha}$
and is also formally conformal invariant.   Its scalar component agrees with that obtained in the preceding subsection.  It is not invariant under the remaining supersymmetries,
as will be discussed below.

\subsection{Connection with the work by Mason and Skinner: superconnection}

In a remarkable recent paper, which appeared while this manuscript was being prepared,
Mason and Skinner \cite{masonskinner} introduced a Wilson loop in momentum twistor space,
whose expectation value they argued (using MHV rules in momentum twistor space \cite{MHVt}) reproduces all-loop scattering amplitudes in $\mathcal{N}=4$ SYM.
The twistor transform of this operator, which they also discussed, gives a Wilson loop in the same four-dimensional space-time
as the Wilson loop we are considering.  This space-time Wilson loop they expressed as the integral over a super-connection,
\be
 \mathcal{W} = \frac{1}{N_c} \mbox{Tr}\left[\mathcal{P} e^{-\int \mathcal{A}}\right]
\ee
where
\ba
 \mathcal{A} &\equiv&  \frac{dx}{2} \left( A + \tilde\psi_A \theta^A + \frac12 D\phi_{AB}\theta^A\theta^B
 - \frac16 \epsilon_{ABCD} D\psi^A\theta^B\theta^C\theta^D
  + \frac1{24}\epsilon_{ABCD} DF \theta^A\theta^B\theta^C\theta^D +\ldots\right)
\nl
&&
+ d\theta^A \left(-\frac12\phi_{AB}\theta^B + \frac13\epsilon_{ABCD} \psi^B\theta^C\theta^D
 + \frac16 \epsilon_{ABCD} F \theta^B\theta^C\theta^D +\ldots\right).  \label{supercon}
\ea
Reference \cite{masonskinner} explained the general method for constructing this operator and gave also explicit expressions up to scalar components.
Here we wrote down a few extra higher-order terms by ``guessing''.  (In each term shown there is a unique way to contract the indices.  The notation is the same as above.)

We believe that the shown terms are still not complete, though, since we also find that, if the object is to transform by super-gauge transformations under supersymmetry,
it should contain other non-linear terms such as $\{\phi_{AC},\phi_{BD}\} (\theta^2)^{AB}(\theta^2)^{CD}$.  The full operator thus presumably contains term going all the way up to order $\theta^8$. 
It would be useful in the future to understand how the twistor transform generates such terms.
 
The two super-Wilson loops look at first sight very different.  One contains corner terms, the other does not.  But this is not an invariant statement: one can always
re-write corner terms as suitable total derivates integrated along edges.  It appears that the superconnection in \cite{masonskinner} achieves this in a particularly clever way: for instance its $\eta_i\eta_i$ component gives
\ba
 \lefteqn{\frac{\eta_i\eta_i }2 \int_0^1 dt \left(
   (1-t)\frac{\tilde\lambda_i \lambda_{i{-}1}D\phi }{2\l i\,i{-}1\r}
   +t \frac{\tilde\lambda_i \lambda_{i{+}1}D\phi }{2\l i\,i{+}1\r}
    + \frac{\l i{-}1\,i{+}1\r\phi}{\l i{-}1\,i\r\l i\,i{+}1\r}
 \right) } && \label{nicescalar} \\
 && =\frac{\eta_i\eta_i }2 \int_0^1dt \left( \frac{\tilde\lambda_i \lambda_{i{-}1}D\phi}{2\l i\,i{-}1\r}
 + \frac{\l i{-}1\,i{+}1\r\phi}{\l i{-}1\,i\r\l i\,i{+}1\r} \frac{d}{dt}(t\phi) \right)
\ea
where we have used the Schouten identity.
The first term is precisely our edge term while and the total derivative gives exactly our vertex term!
We have also checked other scalar components.  The agreement is even more striking
for products like $\eta_i\eta_{i{+}1}$, which are pure vertex terms for us but which appear in Mason and Skinner's edges as pure total derivatives.
Given the role played by supersymmetry in both constructions, we expect complete agreement for the other terms as well.

A second minor difference is that the connection in \cite{masonskinner} contains no factors of the coupling.  Getting rid of factors of the coupling in (\ref{wloop}) is actually easy,
if one rescales $\tilde\psi\to c_0\tilde\psi$, $\phi\to c_0^2\phi$ and $\psi\to c_0^3\psi$.
(This alters the reality condition $(\tilde\psi^\alpha)^*=\psi^\alpha$, but within a Grassman path integral it is not clear whether this has any significance.)
To get rid of the coupling that multiplies the field strength operator one can introduce an auxiliary field $G^{\alpha\beta}$, \emph{\`a la} Chalmers-Siegel \cite{chalmerssiegel} and replace $F/c_0^4\to G$ inside the Wilson loop.  After all this the Lagrangian density (\ref{Lag}) becomes
\ba
 \frac{1}{g^2}\mathcal{L} &\to& \frac{N_c}{4\pi^2} \left( \frac12 F_{\alpha\beta}G^{\alpha\beta}
 + \frac38 (D_{\alpha\dot\alpha}\tilde\psi_A^{\dot\alpha})\psi^{\alpha A}
-\frac18 \tilde\psi_A^{\dot\alpha}D_{\alpha\dot\alpha} \psi^{\alpha A}
 + \frac18 \phi_{AB}  D^2\phi^{AB}
+ \frac14\phi^{AB}[\tilde\psi_{\dot\alpha A},\tilde\psi^{\dot\alpha}_B]
  \right)
 \nl
 && - \frac{g^2N_c^2}{(2\pi)^4} \left( \frac14 G_{\alpha\beta}G^{\alpha\beta}
-\frac14\phi_{AB}[\psi_\alpha^A,\psi^{\alpha B}]
 -\frac1{64}[\phi_{AB},\phi_{CD}] [\phi^{AB},\phi^{CD}] \right).  \label{chalmersiegel}
\ea
An important observation is that the first line is an action for self-dual $\mathcal{N}=4$ \cite{chalmerssiegel}.  In particular, at order $g^0$ (tree-level),
the super-Wilson loop depends only on the self-dual sector.  This is quite a surprise, and was also noted in \cite{masonskinner}.


\subsection{The chiral better-half}
\label{sec:betterhalf}

Tree amplitudes are invariant under both chiral and antichiral supersymmetries.
The above makes manifest the $q_A^\alpha$ and $\tilde s_A^{\dot\alpha}$ supersymmetries,
but what about the other chiral half?\footnote{The author is indebted to E.~Sokatchev for emphasizing this question.}

If one acts with the other half one finds for instance
\be
 \left( -\sum_i \theta_i^{A\alpha}\frac{\partial}{\partial x_i^{\alpha\dot\alpha}} + c_0 \tilde q^A_{\dot\alpha}\right) 2\mathcal{E}_i(t)
 = \tilde\lambda_{i \dot\alpha} ( \lambda_i F\theta^A(t) + \frac12 [\phi_{BC},\phi^{AC}]\eta_i^B + c_0 \lambda_i\psi^A) + \ldots \label{susyvar}
\ee
In particular, this is nonzero.  However, how big is this?
%

If one uses the field rescaling just described, one trivially finds that this operator is explicitly of order $\lambda$. That is, it contributes to $\mathcal{O}(\lambda)$ to any expectation value.
This way the super-Wilson loop makes completely manifest the fact that tree amplitudes are invariant under a larger set of symmetries than loop amplitudes.
Physically, this enhancement must clearly be related to the self-duality property which holds at zeroth order, although it is not yet completely clear how.


The breaking of dual \emph{super}conformal invariance in this formulation is thus explicit and unrelated to regularization.  It exactly parallels what
which was observed at the level of the answer, in the original paper on dual superconformal invariance \cite{sokatchevDCI}.
As an operator, the super-Wilson loop is invariant under only a chiral half of the supersymmetries.

\section{A simple 1-loop NMHV example}
\label{sec:example}

To illustrate the formalism, let us describe a particularly simple helicity component of the 1-loop NMHV amplitude
proportional to $\eta_1\eta_2\eta_3\eta_5$.
It is easy to very from the NMHV tree formula that this component vanishes at tree-level.
This ensures that the one-loop correction is finite.  In fact its value can be extracted from the general results for the NMHV ratio function
that have been obtained in \cite{sokatchevDCI,elvang}, in the scattering amplitude context,
\be
\frac{1}{\l 1235\r} \log u_2 \log \frac{u_3}{u_1}
\ee
where
\be
 u_1= \frac{ \l1234\r\l4561\r}{\l1245\r\l3461\r}, \quad
 u_2= \frac{ \l2345\r\l5612\r}{\l2356\r\l4512\r}, \quad
 u_3= \frac{ \l3456\r\l6123\r}{\l3461\r\l5623\r}.
\ee

In the super Wilson loop, two diagrams contribute to this helicity configuration:
\be
\raisebox{-1.7cm}{\includegraphics[scale=1.0]{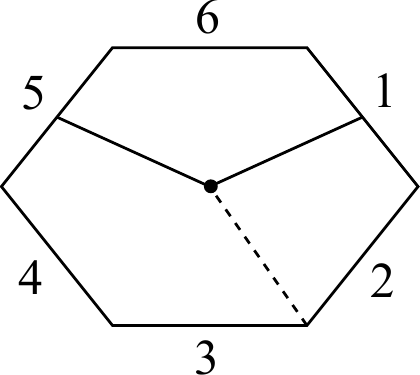}}
\quad+\quad
\raisebox{-1.7cm}{\includegraphics[scale=1.0]{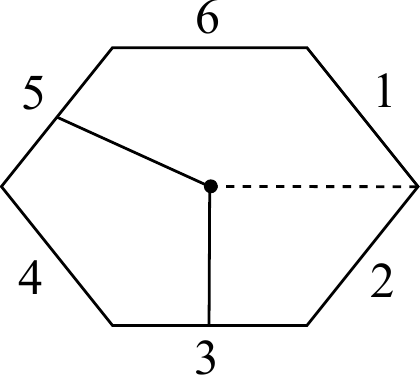}}
\nonumber
\ee
Notice that the same topology contributes to amplitudes at various loop orders depending on which fields enter it:
the shown diagrams contribute to 2-loop MHV amplitudes, 1-loop NMHV amplitudes, and tree N${}^2$MHV amplitudes,
depending on how many powers of $\eta$ are extracted.

We would like to check that these diagrams indeed reproduce the answer known from the scattering amplitude side.
Notice we have a six-dimensional integral: 2 over the edge insertions and 4 over the position of the vertex.
We can perform these integrations in various orders.

For instance we can do the edge integrations first.
The fermion field sourced by edge 5 is
\be
 \psi(y) = \int_0^1 dt \frac{\tilde\lambda_5 (x_5-y)}{(x_4+tp_5-y)^4} \propto \frac{\tilde\lambda_5 (x_5-y)}{(y-x_4)^2(y-x_5)^2}
\ee
and a similar formula gives the fermion fields sourced by edge 1.
Inserting these result into the Yukawa vertex the first diagram gives for instance
\ba
 \int d^4y
 \frac{ [1(y-x_1)(y-x_5)5]}{(y-x_6)^2(y-x_1^2)(y-x_2)^2(y-x_4)^2(y-x_5)^2\l 23\r}.
\ea
This has the form of a $d^4x$ loop integral as typically found in the scattering amplitude context.
With the aim of making conformal invariance manifest we rewrite this as a momentum twistor integral.  Using
the notation of \cite{abcct} the two diagrams just become\footnote{The symbol $\l AB\,i\,j\,k\cap l\,m\,n\r \equiv \l A\,i\,j\,k\r\l B\,l\,m\,n\r - (A\leftrightarrow B)$.}
\be
 \int_{AB}
   \frac{ \l AB\,456\cap612\r\l AB34\r - \l AB\, (456)\cap 234\r \l AB61\r}
   {\l AB12\r\l AB23\r\l AB34\r\l AB45\r\l AB56\r\l AB61\r}
\ee
Notice that the sum of the two diagrams is manifestly finite  (all soft and hard collinear regions are killed by the numerator factor)
and can be thus done directly in four dimensions.  The integral can be done relatively easily using Feynman parameterization,
but this will not discuss it further here, as we will now turn our focus to the more ``Wilson-loop'' form.  Finite integrals
of this type will be discussed at length in a coming publication \cite{inprep}.

In the Wilson loop formulation it is possible to integrate over the vertex first.  Using the three-point function
\be
 \langle \tilde\psi^{\dot\alpha}(x)\tilde\psi^{\dot\beta}(y)\phi(z)\rangle \propto \frac{ (x-z)^{\dot\alpha}{}_{\alpha} (y-z)^{\dot\beta\alpha}}{(x-z)^2(y-z)^2(x-y)^2}.
\ee
the first diagram is thus written as
\be
 \int_0^1 dadb \frac{ [1(x_2-x_1)(x_2-x_5)5]}{(x_4+ap_5-x_2)^2((1-b)p_1+p_2)^2(x_4+ap_5-x_6-bp_1)^2\l23\r}
\ee
Converted to momentum twistor variables the sum of the two diagrams becomes
\be
\l 2456\r \int_0^\infty \frac{d\tau_Ad\tau_B  \l 1235\r }{\l 5A1B\r\l 5A23\r\l 1B23\r_\textrm{reg}}
+
\l 2456\r \int_0^\infty \frac{d\tau_Ad\tau_C  \l 1234\r }{\l 5A12\r\l 5A3C\r\l 123C\r_\textrm{reg}}
\ee
where $Z_B\equiv Z_3+\tau_BZ_5$ and $Z_C\equiv Z_1+\tau_C Z_5$.  The $\tau$ variables are related to the
$a,b$ variables through $a=\frac{\l 56\r\tau_A}{(\l 54\r+\l56\r\tau_A)}$.  The advantage of the $\tau$ parametrization
is that it removes all 2-brackets, thereby making conformal invariance manifest; the $\tau$-contours go from 0 to infinity
with constant argument such that e.g. $\tau_A \frac{\l56\r}{\l54\r}$ is real and positive.

Notice that divergence cancels between the two terms, but the individual
(single-logarithmic) divergences as $\tau_B\to 0$ and $\tau_C\to 0$ have to be regulated using dimensional regularization.
This amounts to $\l 1B23\r \to \l 1B23\r (\frac{\l 1B23\r}{\l 1B\r\l23\r})^{-\epsilon}$.
The integrals are not very difficult and are carried out in Appendix; the result
\be
\frac{1}{\l 1235\r} \log u_2 \log \frac{u_3}{u_1}
\ee
agrees precisely with the one from the scattering amplitude side, confirming the duality at NMHV 1-loop.

This confirms the duality, but also gives a way to
generate many very interesting relations between integrals of very different type.

\section{BCFW recursion relations: tree-level}
\label{sec:tree}

A powerful tool in the computation of scattering amplitudes is the so-called BCFW on-shell recursion relation.
Similar recursion relations should thus apply on the Wilson loop side.
We will now see that this is the case.  In fact, the recursion relations are so powerful that this can be viewed as a proof of the duality.

Actually, the first thing to check is that the Wilson loop at tree-level is a rational function.  We have seen in our tree NMHV example above
how the specific, chiral, form of derivatives on the operators ensures this.

One general argument for rationality
at order $\lambda^0$ goes as follows.  The argument uses the antichiral supersymmetries $\tilde q^A$ and $s^A$, which are valid at this order.
They imply that the Wilson loop at this order is invariant under the full superconformal algebra (32 real supercharges); however ``there is no way to supersymmetrize a logarithm'':
all such invariants are rational functions \cite{sDCI2}.  As will be noted below, it seems conceivable that a much weaker argument based on the fact that we
are computing a tree-level correlation function in a conformal field theory would suffice, but this will not be pursued here.

\subsection{Factorization limits}

Let us thus consider the singularities which arise when two corners of the Wilson loop, say $x_n$ and $x_i$, become null-separated from each other.
Any singularity in this limit will be due to
fields sourced at these corners, or on the edges very close to these corners.  We can use supersymmetry to vastly reduce the number of fields which are sourced.
Indeed, by using twelve of the $Z^\alpha \frac{\partial}{\partial\eta A}$ (which are true symmetries of the Wilson loop)
we can put $\eta_n=\eta_1=\eta_i=0$.  We do not want to try to put $\eta_{i{+}1}=0$ as well,
because the transformation which would achieve this becomes singular when $\langle n\,1\,i\,i{+}1\rangle\to0$, which is the limit we are interested in.  However,
there is no problem in putting three of the $\eta$'s to zero.

Let us be more explicit.  Imagine we want to use supersymmetry to
set $\eta_n$,$\eta_1$, $\eta_i$ and $\eta_k$ to zero, where $k$ is some other particle's label.
Then we can solve explicitly for the shifted $\eta$
\be
 \eta'_i = \frac{ \l n\,1\,i\,k\r \eta_i + \mbox{cyclic}}{\l n\,1\,i\,k\r}.
\ee
These obey $\eta'_n=\eta'_1=\eta'_i=\eta'_{k}=0$, yet the Wilson loop is unchanged $W(\eta')=W(\eta)$.
The choice $k=i+1$ would clearly be singular in the limit considered, but any other choice is perfectly fine.

With the $\eta'$ variables the computation simplifies dramatically.  Only $A_\mu$ is sourced near the corner $x_n$ of the Wilson loop, in Feynman gauge,
\be
 A^\mu(x) \sim \frac{g^2}{8\pi^2} \log (x-x_n)^2 \times \left( \frac{p_n^\mu}{p_n{\cdot}(x-x_n)} - \frac{p_1^\mu}{p_1{\cdot}(x-x_n)}\right).
 \label{inducedfield}
\ee
We are ignoring terms that are nonsingular as $(x-x_n)^2\to 0$.
Unless this hits some derivative there will be no pole in $1/(x_i-x_n)^2$. We need either a field strength operator
inserted at the corner $x_i$, or a derivative of the field strength integrated over edges close to $x_i$.
Since $\eta'_i=0$, using the form (\ref{wloop}), the only term is from $\eta'_{i{+}1}{}^4$ terms integrated along segment $i{+}1$:
\be
 \eta'_{i{+}1}{}^4\int_0 \frac{dt \tilde\lambda_{i_{+}1}\lambda_i\lambda_i\lambda_i }{2 \l i\,i{+}1\r^3} DF(x_i+tp_{i{+}1})
   \sim \frac{g^2}{4\pi^2c_0^4} \frac{\l n1\r\l i\,i{+}1\r}{ \l n\,J\r\l J\,1\r\l i\,J\r\l J\,i{+}1\r} \frac{1}{(x_n-x_i)^2} \frac{\l i\,J\r^4 \eta'_{i{+}1}{}^4}{\l i\,i{+}1\r^4}
   \label{prefactor}
\ee
up to non-singular terms.
The spinor $\lambda_J$ is the one associated with the null vector $(x_i-x_n)$.
We'll simplify this expression in a second.

In general there are interaction vertices in the bulk.  We have to describe how they affect the single propagator contribution we have just computed.
Scalar propagators near the light cone were analyzed in \cite{Aldayetal}, where it was shown
that the singular behavior simply gets dressed by a Wilson line.
The same will be true here.
We cannot directly use their result because we are considering a gauge field instead of a scalar field, however a simple extension will suffice.
Actually we will not need to work very hard because at this stage we only need a tree-level result
(in \cite{Aldayetal}  the all-loop story was worked out). (Loops will be discussed below.)

The argument goes as follows.  First one notes that any interaction vertex with no derivative,
inserted along the propagator joining $x_n$ to $x_i$, will remove the singularity. Thus one needs only consider vertices with a derivative
acting on the singular propagators.  These derivatives are essentially proportional to the (nearly) null vector
$(x_i-x_n)$.  All this is as in \cite{Aldayetal}, here we only have to make sure that the index structure of the gauge three-point vertex works out properly.
Actually this is trivial, because $(x_i{-}x_n)$ vanishes when dotted into the induced field (\ref{inducedfield}), or into the field strength
operator at the other end of the propagator.  Thus $(x_i-x_n)$ must be dotted into the external gauge field, which reduces the vertex to a simple eikonal one.
(The same argument rules out the gauge-scalar-scalar vertex.)
We refer to \cite{Aldayetal} for more details about the computation starting from this point.

Thus the singular propagator (\ref{prefactor}) just gets dressed by an adjoint Wilson line.  This means that the tree-level correlation function
becomes the product of two Wilson loops, in the planar limit, since there are no interactions between the two sides.
\be
\raisebox{-1.8cm}{\includegraphics[scale=1.2]{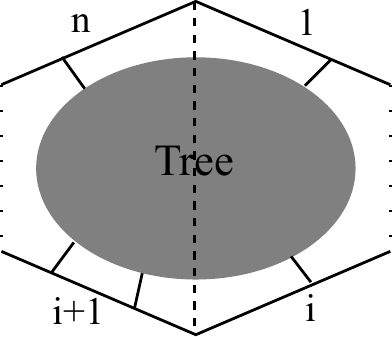}}
\quad\to\quad [n\,1\,i\,i{+}1] \times
\raisebox{-1.8cm}{\includegraphics[scale=1.2]{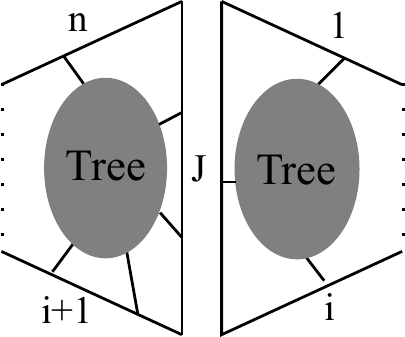}}  \nonumber
\ee

Taking residue $1/(x_n-x_i)^2\to \delta((x_n-x_i)^2)$ for prettier notation and expressing the prefactor (\ref{prefactor}) in terms
of momentum twistors, which is done by simple algebraic manipulations, we thus obtain the singular behavior:
\be
 W_n(1,\ldots,n) \to [n\,1\,i\,i{+}1] \times W_{i{+}1}(1,\ldots,i,J) W_{n-i+1}(J,i{+}1,\ldots,n)
\ee
where $\mathcal{Z}_J$ is the momentum super-twistor
\be
 \mathcal{Z}_J = \l k\,i\,i{+}1\,[n\r \mathcal{Z}_{1]}.
\ee
(On the pole this is independent of $Z_k$.) This is exactly how the tree amplitudes $M_n$ factorize!
This has been derived when $\eta'_n=\eta'_1=\eta'_i=\eta'_{k}=0$.  However, being an equality between manifestly
supersymmetric objects, it holds for arbitrary values of the $\eta_i$.

In this expression
\be
[n\,1\,i\,i{+}1]= \delta(\l n\,1\,i\,i{+}1\r) \frac{\delta^{0|4}(
  \l k\,n\,1\,i\r \eta_{i{+}1}
 +\l k\,1\,i\,i{+}1\r \eta_n
 +\l k\,i\,i{+}1\,n\r \eta_1
 +\l k\,i{+}1\,n\,1\r \eta_i)}
 {\l k\,n\,1\,i\r \l k\,1\,i\,i{+}1\r \l k\,i\,i{+}1\,n\r \l k\,i{+}1\,n\,1\r}.
\ee
is a supersymmetric version of a cut propagator.
(This is also independent of $Z_k$, as follows from the fact that, on the support of the bosonic delta function,
all momentum twistors lie in the same 3-dimensional subspace.)

\subsection{Tree-level BCFW recursion relations}

We have just seen that singularities of the super-Wilson loop factor the same way tree-level scattering amplitudes do.
From this it is a short step to introduce the BCFW recursion relations \cite{BCF,BCFW}
(see also \cite{brandhubernote} for the supersymmetric version).
Let us continue to discuss it here using momentum twistor variables.  One considers the (super-)deformation
of the external data
\be
 \mathcal{\hat Z}_n=\mathcal{Z}_n + z\mathcal{Z}_{n{-}1}
\ee
where $z$ is a complex parameter.  In the Wilson loop picture this corresponds to displacing the point $x_n$ but not the other points.
Because the Wilson loop at leading order is a rational function of the external data, there is no problem
in evaluating it at complex momenta.

Then one writes
\be
 W = \oint \frac{dz}{2\pi i z} W(z)
\ee
where the contour is a small circle around the origin.  We want to deform the contour and express the integral in terms of the poles of
$W(z)$.  The special values of $z$ to consider are:
\begin{itemize}
 \item
   $z= \infty$.  In this limit $x_n\to x_{n{-}1}$ and the Wilson loop reduces to one with one less particle and $\mathcal{Z}_n$ omitted.
 \item $z= -\frac{\l n\,1\,i\,i{+}1\r}{\l n{-}1\,1\,i\,i{+}1\r}$.  These are the factorization channels quoted previously.
 \item $z= -\frac{\l n\,1\r}{\l n{-}1\,1\r}$.  At this point $x_n$ approaches spacetime infinity. The Wilson loop does not diverge in that limit,
  so there is no pole there.
\end{itemize}
Summing up the residues one thus find
\ba
 W_n(1,\ldots,n) &=& W_{n{-}1}(1,\ldots,n{-}1) 
 \nl && + \sum_{i=2}^{n-3} [n{-}1\,n\,1\,i\,i{+}1]
     W_{i+1}(1,\ldots,i,J_i) W_{n-i+1}(J_i,i+1,\ldots,n-1,\hat n_i)
\ea
where the arguments are momentum super-twistors with
\be
 \hat n_{i} = \l 1\,i\,i{+}1\,[n \r\, n{-}1], \quad
   J_i = \l n{-}1\,n\,1\,[i\r\,i{+}1].
\ee

This is the same recursion relation as obeyed by $M_n$ (see \cite{drummondhenn} for its manifestly dual conformal invariant formulation and \cite{abcct} for its version in momentum twistor space similar to here).
It is valid at zeroth order in the coupling, corresponding to tree-level in the scattering amplitude side.  At this order, $W_4=M_4=1$. Thus $W_n=M_n$ for all $n$.
The super-Wilson loop computes all tree amplitudes of the theory!

\section{Loops from chiral Lagrangian insertions}
\label{sec:loop}

We now discuss loop corrections to the Wilson loop.  It will be very useful to use the Lagrangian insertion procedure used recently by Eden, Korchemsky and Sokatchev
\cite{EKS1,EKS2} in this context.  As we will see, this allows the loop corrections to be matched on each side already at the level of the integrand.
The expectation value of the Wilson loop can be written as the path integral
\be
  \int [dAd\psi d\phi] e^{ i \frac{1}{g^2} \int d^4x \mathcal{L}}W.
\ee
The leading contributions to the super-Wilson loop (dual to tree-level scattering amplitudes) are normalized to be of order $g^0$.
This is ensured by the proper choice for $c_0$.  Thus loop corrections can be obtained directly by taking derivatives with respect to the coupling constant,
which will bring down powers of the Lagrangian.
Actually, this requires some care, since the derivatives will also act on the explicit factors of the coupling in the expression in $W$.
Also, one has to be careful with contact terms in the Lagrangian insertion.

The Lagrangian insertion procedure, in the context of supersymmetric field theories, was first introduced in \cite{intriligator}.
It was used to prove non-renormalization theorems \cite{emerynonrenorm} and also as an efficient tool for 2-loop computations \cite{emery2loops}.
Contact terms play an important role in the consistency of the formalism \cite{emerynonrenorm,contactterms}.

\subsection{Generalities}

Let us first consider a simple example in $\lambda\phi^4$ theory. The expectation value of some operator or product of operators
is written as the path integral (overall normalization omitted)
\be
 \langle \mathcal{O}[\phi] \rangle \equiv \int [D\phi] e^{i\int_y ( -\frac12 (D\phi)^2 - \frac{1}{4!}\lambda\phi^4)} \mathcal{O}[\phi].
\ee
In this form one clearly has the expansion
\be
 \langle \mathcal{O}[\phi]\rangle = \sum_\ell (-i\lambda)^\ell \int_{y_1,\ldots,y_\ell} \left[ \frac{1}{\ell!} \langle \mathcal{O}[\phi] \frac{\phi^4(y_1)}{4!}\cdots \frac{\phi^4(y_\ell)}{4!}\rangle\right]_0
 \label{defperturb}
\ee
where the bracket is to be evaluated to zeroth order in the coupling.
This is just the usual perturbation theory.
Now let us rescale $\phi\to \frac{\phi}{\sqrt{\lambda}}$ to bring the Lagrangian into a form closer to what we have to deal with in a gauge theory.
The same expectation is now written
\be
\int [D\phi] e^{\frac{i}{\lambda} \int_y \mathcal{L}_{\phi^4}} \mathcal{O}[\frac{\phi}{\sqrt{\lambda}}].
\ee
where
\be
 {\mathcal L}_{\phi^4}= \frac12 \phi\partial^2\phi - \frac1{4!} \phi^4.
\ee
The reason for writing the kinetic term in this particular fashion will become clear shortly.

The answer being the same as above we must still be able to compute loops by taking derivatives with respect to the coupling.
But now one also gets terms from the explicit dependence on $\lambda$ of the operator, as well as
from the kinetic term in the Lagrangian.  The equivalence with (\ref{defperturb}) would appear to be lost.
Consider for instance the first derivative of the two-point function (properly normalized so that the tree-level contribution
is of order $\lambda^0$):
\be
 \frac{d}{d\lambda} \langle \frac{\phi(x_1)}{\sqrt{\lambda}} \frac{\phi(x_2)}{\sqrt{\lambda}} \rangle
 = - \langle \frac{\phi(x_1)\phi(x_2)}{\lambda^2} \rangle - i\int d^4y \langle \frac{ \phi(x_1)\phi(x_2)\mathcal{L}(z)}{\lambda^3} \rangle. 
\ee
For a one-loop computation this is to be evaluated to order $\lambda^0$.
Now the first term would appear to be of order $\lambda^{-1}$, however, the second term
contains contact terms since $\langle (D^2\phi(y) -\frac{1}{6}\phi^3(y)) \phi(x)\rangle= i\lambda\delta(x-y)$.
These two sorts of terms are easily seen to exactly cancel each other.
Furthermore, this use of the equations of motion flip the sign
of the potential term, recovering precise agreement with the 1-loop term in (\ref{defperturb}).

The same works at higher-loops.  Derivatives of the explicit $1/\lambda^2$ factors from previous Lagrangian insertions get cancelled
by contact terms where multiple Lagrangian insertions go on top of each other.

The moral is that, in this theory, by writing the kinetic term in the appropriate fashion it is possible to cancel contact terms against
explicit coupling constant factors.  Then it is correct to freely use the equations of motion inside the Lagrangian insertions.

\subsection{Chiral Lagrangian insertions}

All this carry over to any gauge theory, and so we return to $\mathcal{N}=4$ super-Yang-Mills for application.

The Wilson loop (\ref{wloop}) depends explicitly on the coupling constant through the factors $c_0\sim \lambda^{1/4}$.  We would like
to cancel this dependence against contact terms.  This can be canonically achieved by writing $3/4$ of the fermion kinetic term
in the form $(D\tilde\psi)\psi$, and the remaining $1/4$ in the form $-\tilde\psi(D\psi)$.  The precise formula is in Appendix (\ref{Lag}).
Similarly, using the self-dual form for the gauge kinetic term and the proper form for the scalar kinetic term given there, one finds that all contact terms in the Lagrangian cancel
one to one against the explicit coupling dependence of $W_n$. 

Thus the following expansion is valid
\be
\langle W_n \rangle =
 \sum_\ell (i\lambda)^\ell \int_{y_1,\ldots,y_\ell} \left[\frac{1}{\ell!} \langle W_n
   \frac{\mathcal{L}_\textrm{on-shell}(y_1)}{\lambda^2}
   \cdots
    \frac{\mathcal{L}_\textrm{on-shell}(y_\ell)}{\lambda^2} \rangle\right]_0
 \label{insertion}
\ee
where the bracket is to be evaluated to tree-level ($\lambda^0$).
The on-shell Lagrangian is to be evaluated from (\ref{Lag}) by freely using the equations of motion, dropping all contact terms.  This gives
\be
 \mathcal{L}_\textrm{on-shell} = \frac14 F_{\alpha\beta}F^{\alpha\beta} -\frac14\phi_{AB}[\psi_\alpha^A,\psi^{\alpha B}]
- \frac1{64}[\phi_{AB},\phi_{CD}] [\phi^{AB},\phi^{CD}].   \label{Lonshell}
\ee
Remarkably, this also happen to be a \emph{chiral} operator: $q_A^\alpha \mathcal{L}_\textrm{on-shell}=0$, \emph{without any total derivative}, as is easily verified.\footnote{
 This chirality property also ensures that contact terms associated with coincident Lagrangian insertions produces terms proportional
  to $\mathcal{L}_\textrm{on-shell}$, guaranteeing the validity of the formula (\ref{insertion}) to all loops.}.

This is also obviously a conformal primary of dimension 4
(this especially obvious since, by virtue of the Lagrangian insertion procedure, it is only ever needed to tree-level order).
Thus it is also invariant under the superconformal generators $\tilde s_A^{\dot\alpha}$.  Under the remaining half of the supersymmetries
$\tilde q^A_{\dot\alpha}$ and $s^A_\alpha$ it is not invariant, but transforms by total derivatives, for instance
\be
 \tilde q^A_{\dot\alpha} \mathcal{L}_\textrm{on-shell} = -\frac12 D_{\dot\alpha\alpha} \left( \psi_{\beta}^A F^{\alpha\beta}
  +\frac12 \psi^{B\alpha} [\phi^{AC},\phi_{BC}]
 \right). \label{anomaly}
\ee

We note that this is not quite the same as the Lagrangian insertion used in \cite{EKS1} or in \cite{EKS2}. It is however
very simply related to, for instance, (3.16) in \cite{Aldayetal}.  One only has to rescale the left-handed and right-handed fermions separately in (3.12)
and remove the anti self-dual part of the field strength by adding $\theta$-term.
Complete equivalence of the results is therefore expected.\footnote{Our form seems also to have been used in \cite{howe}, as the top component of the superfield
used there.  I thank E.~Sokatchev for this observation.}
We insist, however, that the form (\ref{insertion}) is essentially forced upon us by the coupling constant dependence of the super Wilson loop.
Any other writing of the kinetic terms would lead to uncanceled contact terms.

These cancellations are suggestive that a field redefinition can remove the coupling dependence in $\mathcal{W}$ altogether,
as in the above $\phi^4$ example.
In fact, this field redefinition is the Chalmers-Siegel one already mentioned around (\ref{chalmersiegel}).  We could have derived the above
expansion quite directly starting from that form.

\subsection{BCFW for the loop integrand}

The first important remark is that the Lagrangian insertion procedure yields an integrand that is a \emph{rational} function of its
arguments $y_j$.  This was found in explicit computations \cite{EKS1}.  In that context this was more or less guaranteed by the formulation
as a tree-level correlation function in a conformal field theory.

In our $\mathcal{O}(\lambda^0)$ analysis above, we used dual superconformal invariance to
argue for rationality, but it appears a bit more difficult to make such an argument rigorous here since the Lagrangian insertion is invariant only up to a total derivative
(although it is expected physically that this is always the total derivative of a rational function, as found in the scattering amplitude side \cite{abcct}.
 If this could be proved directly the argument would thus carry over.)
In \cite{EKS1}, the main driving factor for rationality seemed to have been that the integrand was identified with a tree-level correlation function in a conformal 
field theory; thus it is also conceivable that a proof based on the weaker condition of conformal invariance is possible.

This being said, it appears very likely that the integrand defined by Lagrangian insertion is a rational function in general.
In what follows we will assume that this is the case, which will be justified \emph{a posteriori} by the remarkable self-consistency of the emerging picture.
So we let the integrand be a rational function of the insertion points and external data.
Then we can continue using the BCFW deformation technology we used at tree-level to determine it.

The contributions discussed in the previous section all go unscathed: they produce products of lower-point amplitudes, only, now
one has to distribute the Lagrangian insertions in all possible ways between the two sides.
However one expects new contributions too: terms where the deformed vertex $\hat{x}_n$ becomes null-separated from a Lagrangian insertion point.
In the scattering amplitude context \cite{abcct} these terms are called single-cuts and they coincide with the forward limit of lower-loop,
higher-point scattering amplitudes.
\be
 \raisebox{-1.6cm}{\includegraphics[scale=1.0]{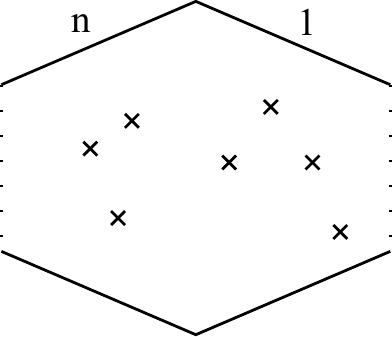}}     \quad\to\quad\sum\quad
 \raisebox{-1.6cm}{\includegraphics[scale=1.0]{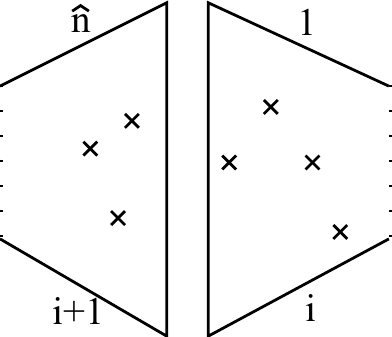}}   \quad+\quad\sum\quad
 \raisebox{-1.6cm}{\includegraphics[scale=1.0]{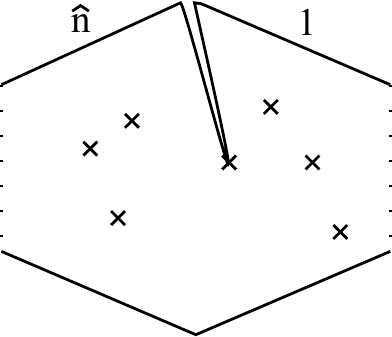}}
 \nonumber.
\ee

These ``single-cut'' terms were in fact our motivation for defining a super-Wilson loop, so that it could reproduce those terms.
Let us thus conclude our story by verifying that the singularities of Lagrangian insertions indeed coincide with forward limits of $W_{n{+}2}$.

\subsubsection{Single-cuts}
\label{sec:singlecut}

This computation can again be greatly simplified if one judiciously uses supersymmetry to set
$\eta_{n{-}1}=\eta_n=\eta_1=0$.  In that case the fields which are singular at null separation from the corner $(n1)$ are simply (\ref{inducedfield}).
Only the $F^2(y)$ term in a Lagrangian insertion will be singular, giving
\be
 \frac{g^2}{16\pi^2} \frac{1}{x_n-y)^2} \frac{\l n\,1\r \lambda_{\hat{A}}\lambda_{\hat{A}}}{\l n\,\hat{A}\r\l \hat{A}\,1\r} F(y).
\ee
Interactions dress this propagator by an adjoint Wilson line, as verified above.
Thus we have a fundamental Wilson loop with two more edges and a field strength inserted at the new corner, as claimed in the Introduction.

In the BCFW formula the corresponding pole adds a new term which is a Wilson loop with a shifted point $\hat x_n$
connected by an adjoint Wilson line to the insertion
\be
 \frac{g^2}{16\pi^2} \frac{1}{\l AB\,n{-}1\,n\r\l AB\,n\,1\r\ AB\,n{-}1\,1\r\l AB\r^3}
 \times \lambda_{\hat{A}}\lambda_{\hat{A}} F(y).
\label{BCFWcut}
\ee
Here the line $(AB)$ corresponds to the point $y$ and we have multiplied by a trivial factor $1/\l AB\r^4$ for future convenience
(this factor enters the Jacobian $\int d^4y \to d^4 Z_AZ_B/\l AB\r^4$) and
\be
 \hat n = \l AB\,1\,[n{-}1\r\, n], \quad \hat A= \l n{-}1\,n\,1\,[A\r\, B].
\ee
This normalization for $\hat{A}$ is important here since (\ref{BCFWcut}) is not homogeneous in it.
This computation is valid when $\eta_{n{-}1}=\eta_n=\eta_1=0$.

We have to compare this result with the forward limit of the super-Wilson loop (\ref{wloop}) with $n+2$ particles.

In the scattering amplitude context the forward limit is a somewhat singular limit but it is well-defined, for massless particles, at least
in any theory with $\mathcal{N}=1$ supersymmetry \cite{loopandtrees}.  A form particularly convenient for $\mathcal{N}=4$ was given recently in \cite{abcct}.
This form uses a contour integral to express the result in terms of strictly non-forward, hence manifestly well-defined, data.
It looks even prettier when inserted into the BCFW formula, giving rise the BCFW term \cite{abcct}
\be
 \int d^4\eta_A d^4\eta_B \oint_{\textrm{GL(2)}} [A'B'\,n{-}1\,n\,1] M_{n{+}2} (1,\ldots,\hat n, \hat A, B').
\ee
The contour integral is over GL(2) matrices $G$ sending $(A,B)$ to $(A',B')=(A,B)G$, and is to be done on residues around $A'\propto B'\propto \hat A$.

So one expects that the above expression (\ref{BCFWcut}) is equal to
\be
 \int d^4\eta_A d^4\eta_B \oint_{\textrm{GL(2)}} [A'B'\,n{-}1\,n\,1] W_{n{+}2} (1,\ldots,\hat n, \hat A, B')  \label{good_forward_limit}.
\ee
The verification is a straightforward computation using the explicit form of the edge and vertex terms.  This computation is reported
in Appendix with the result that expressions (\ref{BCFWcut}) and (\ref{good_forward_limit}) agree precisely.

\subsubsection{Recursion relations}

This equality allows us to write the BCFW formula for the loop
integrand defined by Lagrangian insertion in a form similar to that in \cite{abcct}:
\begin{eqnarray}
 W_n(1,\ldots,n; \{AB\}_\ell) &=&
 W_{n{-}1}(1,\ldots,n{-}1; \{AB\}_\ell) \nl
 &&  \hspace{-3cm}
 + \frac{1}{\ell!}\sum_{\sigma_\ell} \sum_{i=2}^{n{-}3} [i\,i{+}1\,n{-}1\,n\,1] ~
   W_{i{+}1} (1,\ldots,i,J_i;\{ AB\}_L)
   W_{n{-}i{+}1} (J_i,i{+}1,\ldots,\hat{n}; \{AB)\}_R)
  \nl
  && \hspace{-3cm}
   +\frac{1}{\ell} \sum_j \oint_{\textrm{GL(2)}} [A_jB_j\,n{+}1\,n\,1] ~
   W_{n{+}2} (1,\ldots,\hat n_{AB_j},\hat A_j,B_j; \{ AB\}_{\ell \setminus j})   \label{loop_level_BCFW}
\end{eqnarray}
where the shifted momentum (super-)twistors that enter are
\ba
 \hat n_{j} &=& \l i\,i{+}1\,1\,[n{-}1\r\, n], \quad   J_i = \l n{-}1\,n\,1\,[i\r\,i{+}1], \nonumber \\
 \hat n_{AB_j} &=& \l A_jB_j\,1\,[n{-}1\r\, n], \quad \hat A_j= \l n{-}1\,n\,1\,[A_j\r\, B_j].
\ea
In the second line the sum is over all ways of distributing the loop momenta $\{ AB\}_\ell$ into two sets $\{ AB\}_{L,R}$.

This is exactly recursion relation for the loop integrand found in \cite{abcct} in the scattering amplitude context.
It follows that, to all loop orders and for arbitrary external polarizations, the loop integrands for the super-Wilson loop and that for scattering
amplitudes are \emph{identical}.

Both integrals are actually divergent.  The Wilson loop is ultraviolet-divergent while scattering amplitudes are infrared-divergent.
The leading divergences on both side agree  and are both controlled by the so-called
``cusp anomalous dimension:'' this quantity governs the singularities of null Wilson loops as well as the Sudakov factors in scattering amplitudes.
On the other hand, subleading and finite parts need not agree, because natural regularization schemes on the two sides of the correspondence are different.
To finish the proof of the correspondence, it is important to make sure that the logarithm of the amplitude (the relevant quantity)
has a well-controlled scheme dependence.  A general argument for this is presented in Appendix.

\section{Conclusions}
\label{sec:conc}

We have constructed, directly from field theory, a family of supersymmetric Wilson loop operators in $\mathcal{N}=4$ Yang-Mills theory.
This family depends on the same momentum and polarization
data as scattering amplitudes do.  By analyzing the factorization properties of this object in various channels at zeroth order in the coupling,
we have derived recursion relations for it. These recursions are the same as those obeyed by tree-level scattering amplitudes, showing that the two objects are the same.
Furthermore, similar recursion relations extend to appropriately defined ``loop integrands'' which can be defined on each side.
The duality between the two objects thus extends to all loops, already at the level of integrands.

The super-Wilson loop sheds some light on the nature of the duality.  Understanding this at tree level is already nontrivial and interesting,
since we have seen that in a sense loops are made by sewing trees.
In the Wilson-loop side it turns out that only the self-dual sector contributes to tree amplitudes (this is especially clear in
the Chalmers-Siegel form (\ref{chalmersiegel}) and even more so in the work \cite{masonskinner}).
This immediately makes a striking prediction on the scattering amplitude side: \emph{there must be a subsector
of $\mathcal{N}=4$ SYM that is sufficient to compute all tree amplitudes.}  
This sector is certainly not directly the self-dual sector, because the asymptotic fields of negative helicity gluons are not self-dual by definition.
However it should be somehow close to the self-dual sector.
The amount of its non-self-duality is presumably related, through the T-duality, to the non-self-duality of the fields on the support of the Wilson loop.
So at this stage it is not clear what the subsector is, and what its relation to self-dual $\mathcal{N}=4$ is, but
it might be productive to understand this issue.

It is important to note that we have not computed completely the form of the Wilson loop operator: we have computed only those components that we needed,
dropping certain nonlinear terms containing with 5 or more powers of $\theta$ at the corners.
Most of our results regarding factorization and recursion relations assumed little more than the existence of
a supersymmetry-covariant Wilson loop, and supersymmetry always allowed us to make do without these terms.
Nevertheless, since to our knowledge such a superconnection has never be written down in complete detail, we feel that it would
be good to see it worked out fully somewhere.

On a technical side, the examples that we have worked out have the unsatisfactory feature that conformal invariance is not
manifest at all intermediate steps.  The six-dimensional formalism for four-dimensional conformal field theories, presented recently
by Weinberg \cite{weinberg}, might be very useful in that respect.
Especially when combined with momentum twistors, which are in the spinor representation of SO(4,2) and allow for particularly compact expressions for
chiral conformal invariants, this should provide a particularly efficient formalism.

Another question on which the super-Wilson loop sheds some light is the longstanding observation that
tree amplitudes are invariant under a full Yangian symmetry, while loop amplitudes  (even those which are finite)
do not exhibit it \cite{sokatchevDCI}.  In particular, half the dual supersymmetries $d/d\theta$ are preserved
to all loops, while the other half $\theta d/dx$ is broken starting from 1-loop.  This has been understood as a holomorphic anomaly in \cite{sokatchevsDCI}.
The Wilson loop defined here and in \cite{masonskinner} makes the unbroken half of the supersymmetry fully manifest but explicitly breaks the other half,
As discussed in section (\ref{sec:betterhalf}), the Wilson loop, constructed from various other requirements, turns out to violate explicitly
these symmetries but by an amount explicitly proportional to $g^2$.  Thus the empirical observation of \cite{sokatchevDCI} are built-in into the operator.
The symmetry enhancement that occurs at tree-level must clearly be related to the self-dual sector,
although precisely how is not completely clear yet.

The explicit breaking suggests a novel way of computing what has appeared as an ``anomaly'' before.
Indeed, we have found that the variation is given by (\ref{anomaly})
\be
 \theta^{A \alpha} \frac{\partial}{\partial x^{\alpha\dot\alpha}} \log \langle W_n\rangle
  = \oint dx_{\dot\alpha\alpha}  \frac{\langle (c_0\psi^A + F\theta^A+ \ldots)^\alpha W_n\rangle}
  {\langle W_n \rangle},
\ee
where the (color-adjoint) parenthesis is to be inserted along the Wilson loop; a similar expression holds for the superconformal generators $s^A_\alpha$. 
As just remarked, the right-hand side is manifestly of order $g^2$.
Furthermore, at any loop order, the right-hand side can have at most single-logarithmic divergences.  In fact, it appears very likely
that these divergences are concentrated to the corners of the $dx$ integral, so that the correlation function itself is finite to any loop order.
If this formula is correct, it might prove a particularly powerful way of computing derivatives of the logarithm of MHV amplitudes at higher-loop levels;
these derivatives, and the full answer, are known to exhibit remarkable simplicity and structure \cite{volovich}.

The author would like to thank Nima Arkani-Hamed, Lionel Mason and David Skinner for useful discussions and Emery Sokatchev
and Gregory Korchemsky for very useful correspondence
and for a careful reading of this draft.
Hospitality is gratefully acknowledged from the Perimeter Institute in Waterloo, Canada, for a wonderful stay during which part of this work was conducted.
The author gratefully acknowledges support from NSF grant PHY-0969448.

\pagebreak

\begin{appendix}

\section{Lagrangian and supersymmetry generators}

We now describe, mostly for the sake of fixing our conventions,
the well-known $\mathcal{N}=4$ Lagrangian in a form which will be convenient for us:
\ba
 \mathcal{L} &=& \frac14 F_{\alpha\beta}F^{\alpha\beta} +  \left[
\frac38 (D_{\alpha\dot\alpha}\tilde\psi_A^{\dot\alpha})\psi^{\alpha A}
-\frac18 \tilde\psi_A^{\dot\alpha}D_{\alpha\dot\alpha} \psi^{\alpha A}
\right]
 + \frac18 \phi_{AB}  D^2\phi^{AB}
 \nl
 &&
+\frac14\phi_{AB}[\psi_\alpha^A,\psi^{\alpha B}]
+ \frac14\phi^{AB}[\tilde\psi_{\dot\alpha A},\tilde\psi^{\dot\alpha}_B]
 + \frac1{64}[\phi_{AB},\phi_{CD}] [\phi^{AB},\phi^{CD}]. \label{Lag}
\ea
This enters the path integral through the factor $e^{ -i \frac{1}{g^2} \int d^4x \mathcal{L}}$.
The self-dual field strength tensor is defined through
\be
 F_{\alpha\dot\alpha\,\beta\dot\beta} = \epsilon_{\dot\alpha\dot\beta} F_{\alpha\beta} +
 \epsilon_{\alpha\beta} F_{\dot\alpha\dot\beta}
\ee
where $F_{\mu\nu}= \partial_\mu A_\nu-\partial_\nu A_\mu + [A_\mu,A_\nu]$.
(We suppress all color indices;  $[A_\mu,A_\nu]^a\equiv f^{abc} A_\mu^b A_\nu^c$ and $F_{\alpha\beta}F^{\alpha\beta}\equiv F_{\alpha\beta}^a F^{\alpha\beta a}$.)


The supersymmetry transformations, under which $\mathcal{L}$ transforms by a total derivative, are
\begin{align}
q_A^\alpha A^{\beta\dot\beta}&= \epsilon^{\alpha\beta} \tilde\psi_A^{\dot\beta},
& q_A^\alpha \tilde\psi_B^{\dot\beta} &= D^{\alpha\dot\beta} \phi_{AB}, \\
q_A^\alpha \phi^{BC} &= \psi^{\alpha[B} \delta_A^{C]},
& q_A^\alpha \psi^{B \beta} &= \delta_A^B F^{\alpha\beta} + \frac12\epsilon^{\alpha\beta} [\phi_{AC},\phi^{BC}].
\end{align}
together with their parity conjugates, which are simply
\begin{align}
\tilde q^{A\dot\alpha} A^{\beta\dot\beta}&= \epsilon^{\dot\alpha\dot\beta} \tilde\psi^{A \beta},
& \tilde q^{A\dot\alpha} \psi^{B \beta} &= D^{\dot\alpha\beta} \phi^{AB}, \\
\tilde q^{A\dot\alpha} \phi_{BC} &= \psi_{\alpha[B} \delta^A_{C]},
& \tilde q^{A\dot\alpha} \tilde\psi_B^{\dot\beta} &= \delta^A_B F^{\dot\alpha\dot\beta} + \frac12\epsilon^{\dot\alpha\dot\beta} [\phi^{AC},\phi_{BC}].
\end{align}

Our writing of the kinetic terms in the Lagrangian density may appear somewhat unusual; the motivation for it is explained in the main text.

Let us finally summarize the rest of our notations.
We have $p^{\alpha\beta}\equiv p^\mu \sigma_\mu^{\alpha\beta}$ where $\sigma_\mu^{\alpha\dot\alpha}=(i,-i\vec\sigma) = -\sigma^\mu_{\dot\alpha\alpha}$.
The spinor products are $\langle vw\rangle = v_\alpha v^\alpha$ and $[vw]=v_{\dot\alpha}v^{\dot\alpha}$
where indices are lowered using $v_\alpha=v^\beta \epsilon_{\beta\alpha}$ and $v_{\dot\alpha} = v^{\dot\beta} \epsilon_{\dot\beta\dot\alpha}$ with $\epsilon$
 the antisymmetric symbol with $\epsilon_{12}=\epsilon_{\dot1\dot2}=1$.
Indices are raised using the inverse symbol with $\epsilon^{12}=\epsilon^{\dot1\dot2}=-1$.
The fields obey the reality conditions $\phi_{AB}^*=\phi^{BA}$ and $(\psi^{A\alpha})^\dagger=\tilde\psi_A^{\dot\alpha}$.

\section{Vertex terms}

In this section we report our computation of the vertex term in the super-Wilson loop (\ref{wloop}); for completeness we reproduce
the edge term also.
\ba \hspace{-0.5cm}
 2\mathcal{E}_i &=& -\tilde\lambda_i \lambda_i A
 + \frac1{c_0} \tilde\lambda_i \tilde\psi_A \eta_i^A
 + \frac1{2c_0^2} \frac{\tilde \lambda_i \lambda_{i{-}1} D \phi_{AB}}{\l i\,i{-}1\r} \eta_i^A\eta_i^B \nl
 &&
 - \frac{1}{3!c_0^3} \epsilon_{ABCD}\frac{\tilde\lambda_i \lambda_{i{-}1}\lambda_{i{-}1} D\psi^A}{\l i\,i{-}1\r^2} \eta_i^B\eta_i^C \eta_i^D
 - \frac{1}{4!c_0^4} \epsilon_{ABCD}\frac{\tilde\lambda_i \lambda_{i{-}1}\lambda_{i{-}1}\lambda_{i{-}1} DF}{\l i\,i{-}1\r^3}\eta_i^A\eta_i^B\eta_i^C\eta_i^D,
  \label{edgeA}
\ea
\ba
 \mathcal{V}_{i\,i{+}1} &=& 1+ \frac{\phi_{AB}}{c_0^2\l i{+}1\,i\r} \left(\eta_{i{+}1}^A\eta_i^B - \frac12\frac{\l i{-}1\,i{+}1\r}{\l i{-}1\,i\r} \eta_i^A\eta_i^B\right)
    \nl   &&
      + \frac{1}{3!c_0^3} \frac{ \l i{-}1\,i{+}1\r (\l i\,i{-}1\r \lambda_{i{+}1} + \l i\,i{+}1\r \lambda_{i{-}1}) \psi}{\l i{-}1\,i\r^2\l i\,i{+}1\r^2} \eta_i\eta_i\eta_i
       + \frac1{2c_0^3} \frac{\lambda_{i{+}1}\psi}{\l i\,i{+}1\r^2}\eta_i\eta_i\eta_{i{+}1}
 \nl && 
        - \frac1{2c_0^3} \frac{\lambda_i \psi}{\l i\,i{+}1\r^2} \eta_i\eta_{i{+}1}\eta_{i{+}1}
+ \frac1{4.4!} \frac{\l i{-}1\,i{+}1\r^2}{\l i\,i{-}1\r^2 \l i\,i{+}1\r^2} \phi_{AB}\phi^{AB} \eta_i^4
\nl &&
        + \frac1{4!c_0^4} \frac{ \l i{-}1\,i{+}1\r 
         (\l i\,i{-}1\r^2\lambda_{i{+}1}\lambda_{i{+}1}
      + \l i\,i{-}1\r  \l i\,i{+}1\r \lambda_{i{+}1}\lambda_{i{-}1}
      + \l i\,i{+}1\r^2\lambda_{i{-}1}\lambda_{i{-}1})F}
      {\l i\,i{-}1\r^3\l i\,i{+}1\r^3} \eta_i^4
        \nl  &&
        + \mathcal{O}(\eta_i\eta_{i{+}1}\eta^2) + \mathcal{O}(\eta^5). \label{vertex}
\ea
We found it difficult to find a simple organizing principle for the vertex terms.
Fortunately, the omitted terms in the vertices will not be need for the analyses in this paper.
These omitted terms are all proportional to at least one power of $\eta_i$ and one power of $\eta_{i{+}1}$,
and we expect them to form a nontrivial series ranging all the way up to $\eta_i^4\eta_{i{+}1}^4$.

This is ``supersymmetric'' (under half of the supersymmetries) in the sense that edges transform by a gauge transformation
$(q_A^\alpha + c_0 \lambda^\alpha \frac{\partial}{\partial \eta^A}) \mathcal{E}_i=
 (\partial_t - [\mathcal{E}_i, )X^\alpha_{iA}$, where
\begin{equation}
 X_{iA}^\alpha \equiv  \frac{\lambda_{i{-}1}^\alpha}{c_0 \l i\,i{-}1\r} \left(
  \phi_{AB}\eta_i^B
  - \frac{1}{2c_0} \epsilon_{ABCD}\frac{\lambda_{i{-}1}\psi^B}{\l i\,i{-}1\r} \eta_i^C\eta_i^D
  - \frac{1}{3!c_0^2} \epsilon_{ABCD}\frac{\lambda_{i{-}1}\lambda_{i{-}1} F}{\l i\,i{-}1\r^2} \eta_i^B\eta_i^C\eta_i^D\right).
\nonumber
\end{equation}
The vertices transform by the same ``gauge transformation,''
$(q_A^\alpha + c_0 \lambda^\alpha \frac{\partial}{\partial \eta^A}) \mathcal{V}_{i\,i{+}1}+
X_{iA}^\alpha  \mathcal{V}_{i\,i{+}1}
-\mathcal{V}_{i\,i{+}1}X_{i{+}1\,A}^\alpha=0$
so that the Wilson loop (\ref{wloop}) is invariant.

\section{Integral in the 1-loop example}

In this Appendix we consider the integral
\be
\l 2456\r \int_0^\infty \frac{d\tau_Ad\tau_B  \l 1235\r }{\l 5A1B\r\l 5A23\r\l 1B23\r_\textrm{reg}}
+
\l 2456\r \int_0^\infty \frac{d\tau_Ad\tau_C  \l 1234\r }{\l 5A12\r\l 5A3C\r\l 123C\r_\textrm{reg}}
\ee
which arose in our NMHV 1-loop example, with $Z_A=Z_4+\tau_A Z_6$, $Z_B=Z_2+\tau_BZ_1$ and $Z_C=Z_2+\tau_C Z_4$
and the contours as described in the main text.
The (single-logarithmic) divergences as $\tau_B\to 0$ and $\tau_C\to 0$ are regulated using dimensional regularization,
$\l 1B23\r \to \l 1B23\r (\frac{\l 1B23\r}{\l 1B\r\l23\r})^{-\epsilon}$.
The divergences cancel between the two terms.  After doing the $\tau_B$ and $\tau_C$ integrals one is thus left with
a finite integral
\be
\l 2456\r \int_0^\infty \frac{d\tau_A}{\l 5A12\r\l 5A23\r} \log\left(
  \frac{\l 5A12\r\l5A34\r\l 6123\r}{\l 5A61\r\l 5A23\r\l 1234\r}\right).
\ee
To proceed further we rewrite this as a function of cross-ratios by rescaling $\tau_A\to a\frac{\l 5412\r}{\l 5612\r}$, such that the integral becomes
\be
 \frac{1-1/u_2}{\l 1235\r} \int_0^\infty \frac{da}{(1+a)(1+a/u_2)} \log\left(
  \frac{ a(1+a)}{1+a/u_2} \frac{ \l4512\r^2 \l 3456\r\l6123\r}{\l1234\r\l 2345\r\l 4561\r\l 5612\r}\right).
\ee
We now observe that the rational factor has a symmetry under $a\to u_2/a$.  Upon symmetrizing the integrand,
the $a$-dependence inside the logarithm disappears.  Thus the integrand does not produce dilogarithms but only products of logarithms:
\be
 \frac{1-1/u_2}{\l 1235\r} \int_0^\infty \frac{da}{(1+a)(1+a/u_2)} \log\left(
  \frac{ \l4512\r \l 3456\r\l6123\r}{\l1234\r\l 4561\r\l5623\r}\right) = \frac{1}{\l 1235\r} \log u_2 \log \frac{u_3}{u_1}.
\ee

\section{Forward limits}

In this section we check that the contour integral version of the forward limit
\be
 \int d^4\eta_A d^4\eta_B \oint_{\textrm{GL(2)}} [A'B'\,n{-}1\,n\,1] W_{n{+}2} (1,\ldots,\hat n, \hat A, B')  \label{good_forward_limitA}.
\ee
reproduces just (\ref{BCFWcut}).

A simple and correct way to do the integral over the GL(2) is as follows.
Since invariance under the diagonal component of the GL(2) is already manifest, the integral to be done is really over only the off-diagonal components.
This can be done in two steps.  First one integrates over the off-diagonal component $A'=A+\tau_1 B$.  The only dependence on $\tau_1$
is through the R-invariant, which readily gives
\be
 \oint \frac{d\tau_1}{2\pi i} [A'B\,n{-}1\,n\,1] = \frac{\delta^{0|4}(\eta_A\l B\,n\,n{-}1\,1\r + \textrm{cyclic})}{\l AB\,n{-}1\,n\r\l AB\, n\,1\r\l AB\,n{-}1\,1\r \l B\,n{-}1\,n\,1\r^2}.
\ee
To proceed further we restrict to $\eta_{n{-}1}=\eta_n=\eta_1=0$ as in section (\ref{sec:singlecut}).  In this case $\eta_{\hat{A}}=0$ due to the fermion $\delta$-function:
all the $\eta_A,\eta_B$ dependence of $W$ is through $\mathcal{Z}_{B'}= \ZZ_B + tau_2\ZZ_A$.  Thus we are extracting the component of order $\eta_{B'}^4$.
We can use this to do perform the fermion integrals,
\be
 \int d^4\eta_A d^4\eta_B  \frac{\delta^{0|4}(\eta_A\l B\,n\,n{-}1\,1\r + \textrm{cyclic})}{ \l B'\,n{-}1\,n\,1\r^2} \delta^{0|4}(\eta_B + \tau_2\eta_A)
  = \l B'\,n{-}1\,n\,1\r^2.
\ee
The $\tau_2$ integral is to be done on a small circle around the special value where $\l B'\,n{-}1\,n\,1\r=0$. 
Now all we got so far near this point is a double zero.  However some terms in the Wilson loop (\ref{edgeA}) contain denominators $1/\l i\,i{-}1\r\to \l B'\,A\r$,
which produce poles.  The terms in the Wilson loop contain at most three such powers, which is exactly what we need to obtain a finite residue.
Thus we need only keep those terms
with three powers of $1/\l i{-}1 i\r$.  These are
\be
 \frac{1}{c_0^4} \oint \frac{d\tau_2}{2\pi i} \frac{\l B'\,n{-}1\,n\,1\r^2}{\l AB\,n{-}1\,n\r\l AB\, n\,1\r\l AB\,n{-}1\,1\r \l B'\,\hat A\r^3} \frac{\l \hat{A}\,1\r}{\l B'\,1\r} \times
 \left( -\frac12 \int_0^1 dt \tilde \lambda_{\hat{A}} \lambda_{\hat{A}}\lambda_{\hat{A}}\lambda_{\hat{A}} DF(y + t p) 
+ \lambda_{\hat{A}}\lambda_{\hat{A}} F(y+p)\right).
\ee
The edge integral in the parenthesis
is a total derivative, which cancels precisely against the vertex term at one endpoint.  There remains only the other endpoint, located at $y$.
Finally the contour integral can be expressed as
\be
 \frac{g^2N_c}{16\pi^2} \frac{1}{\l AB\,n{-}1\,n\r\l AB\,n\,1\r\ AB\,n{-}1\,1\r\l AB\r^3} \times \lambda_{\hat{A}}\lambda_{\hat{A}} F(y).
\ee
This agrees precisely with (\ref{BCFWcut}), as anticipated in the main text.

More precisely, this equality has been derived for $\eta_{n{-}1}=\eta_n=\eta_1=0$.  However, both the correlation function with Lagrangian
insertions and the expression (\ref{good_forward_limitA}) are manifestly invariant under $Z^\alpha \frac{\partial}{\partial \eta^A}$ supersymmetries.
Thus it follows that they agree for all values of the $\eta$.

\section{Relationships between different infrared regulators}

Several natural regulators are possible for the Wilson loop:
\begin{itemize}
 \item Dimensional regularization to $D=4-2\epsilon$ dimensions.
 \item Dimensional regularization of the measure in the Lagrangian insertion procedure, keeping the integrand in four dimensions \cite{EKS1}.
 \item Multiplication of the Lagrangion insertion integrand by $\prod_i \frac{(y-x_i)^2}{(y-x_i)^2+m^2)}$ (similar to adding masses in the Higgs-inspired scheme of \cite{aldayhenn}).
 \end{itemize}
and similarly several natural regulators are possible for scattering amplitudes:
\begin{itemize}
 \item Dimensional regularization to $D=4+2\epsilon$ dimensions.
 \item Multiplication of the massless, four-dimensional integrand by $\prod_i \frac{(y-x_i)^2}{(y-x_i)^2+m^2)}$ (similar to above).
 \item Higgs regulator \cite{aldayhenn}.
\end{itemize}

None of these regulators will produce the same answer.  The last two differ in whether masses are kept in numerators.
Here we would like to argue that the logarithm of the Wilson loop,
which is the relevant physical quantity, can only differ from one approach to the other in a very mild and controlled way.

More precisely, the logarithm of the Wilson loop takes the form (see for instance \cite{4ptDCI} and references therein)
\be
 \log \langle W_n\rangle = \sum_i \left( \gamma_\textrm{cusp} \log^2\frac{p_i{\cdot}p_{i{+}1}}{m^2}
+ d \log \frac{p_i{\cdot}p_{i{+}1}}{m^2}
 +C\right) + \mbox{Finite}.
\ee
In the scattering amplitude side this form follows from factorization theorems in the planar limit.

It is completely expected, and verified in practice (to four loops in \cite{henn4loop}), that different regulators only affect the scheme dependent constants
$d$ and $C$ but the ``finite part'' is unchanged.  This is part of the physical content of factorization.
The modest point we would like to make in this Appendix is that factorization is made particularly explicit by the Lagrangian insertion procedure.

First we observe that all the above regulators are compatible with the Lagrangian procedure
(in fact, some are only defined within it).  So, we can use it in our analysis.  The main observation then is that in this procedure
the logarithm can be taken already at the level of the integrand.  That is, for instance, at two-loops, one can consider the integral
\be
 \log \l W_n\r = \int_{y_1,y_2} \left( M^\textrm{MHV}_n(y_1,y_2) -  \frac12 M^\textrm{MHV}_n(y_1) M^{\textrm{MHV}}(y_2)\right).
\ee
The claim is that the parenthesis is much better behaved then each term in it object,
with divergent regions with $y_1$ and $y_2$ not close canceling explicitly.  This can be easily verified at two-loop using the well-known two-loop integrand
for four particles \cite{bern2loop}. More generally, we would like to argue that divergences in the logarithmic occur only from the regions where all 
particles collectively approach the same singular region. In other words, that factorization occurs already at the integrand level.

The mechanism we suggest is very simple:  Lagrangian insertions are color singlet operators.  Imagine that, at $\ell$ loops, a subgroup of
Lagrangian insertions approaches a singular region. There is a scaling parameter $s$, for instance near a hard collinear singular region,
$(y_i^+,y_i^-,y_i^\perp)\to (y_i^+,s^2y_i^-, sy_i^\perp)$, which we can introduce to make all these insertions approach the hard collinear region at the same rate.
Similar scaling variables can be introduced to deal with the soft wide-angle and soft-collinear regions.
In the scaling limit we have a number of color-singlet operators inserted very close to an edge, plus other ``passive'' insertions at some other fixed locations.
We know from general power-counting that the singularity in this limit will be at most logarithmic: the correlation function times the measure factor cannot not blow up in the limit.
The only terms which survive thus behave like dimension-0 color-single corrections to the edge operator, as seen from
the passive insertions.  But such a correction can only be proportional to the edge itself.  That is the field induced by the edge region is not affected, in the scaling limit,
by the presence of the insertions.  This implies the factorization of the integrand in this limit into a product of two correlation functions,
one involving the singular subgroup, and one involving the passive subgroup.

Since such products are killed by the logarithm, we conclude that the integrand for the logarithm of the $W$ is regular whenever a proper subgroup
of insertions approaches a singular region.  As a corollary, if we take the scaling limit as all Lagrangian insertions approach a singular region,
and strip off a collective $d^4y$, the integral over the relative coordinates must be finite.  This integrand multiplies
the zeroth order expectation value of the Wilson loop.  Since the mechanism is the color neutrality of the Lagrangian insertion, it applies in $D$ dimensions as well
and thus implies the equality of the first three regulators listed above, up to the finite ``constants under the logarithm''.
(The scaling limit in the soft directions, under which soft singularities factorize independently, is multiplicative.
Thus the integral over relative rapidities, of the integrand for hard collinear divergences, is guaranteed to be finite.)

A definite prediction of this argument is that there should exist a finite $4(\ell-1)$-dimensional integral which gives the cusp anomalous dimension to $\ell$ loops.
Its integrand is the connected correlation function of $\ell$ Lagrangian insertions in the universal background field very close to a segment of the Wilson loop,
that is sourced by this segment and the two cusps at its endpoints.

\end{appendix}


\begin{thebibliography}{99}

\bibitem{minahan}
  J.~A.~Minahan and K.~Zarembo,
  ``The Bethe-ansatz for N = 4 super Yang-Mills,''
  JHEP {\bf 0303}, 013 (2003)
  [arXiv:hep-th/0212208].
\bibitem{BES}
  N.~Beisert, B.~Eden and M.~Staudacher,
  ``Transcendentality and crossing,''
  J.\ Stat.\ Mech.\  {\bf 0701}, P021 (2007)
  [arXiv:hep-th/0610251].
\bibitem{maldacena}
  J.~M.~Maldacena,
  Adv.\ Theor.\ Math.\ Phys.\  {\bf 2}, 231 (1998)
  [Int.\ J.\ Theor.\ Phys.\  {\bf 38}, 1113 (1999)]
  [arXiv:hep-th/9711200].

\bibitem{green}
  M.~B.~Green, J.~H.~Schwarz and L.~Brink,
  ``N=4 Yang-Mills And N=8 Supergravity As Limits Of String Theories,''
  Nucl.\ Phys.\  B {\bf 198}, 474 (1982).




\bibitem{4ptDCI}
  Z.~Bern, M.~Czakon, L.~J.~Dixon, D.~A.~Kosower and V.~A.~Smirnov,
  ``The Four-Loop Planar Amplitude and Cusp Anomalous Dimension in Maximally
  Supersymmetric Yang-Mills Theory,''
  Phys.\ Rev.\  D {\bf 75}, 085010 (2007)
  [arXiv:hep-th/0610248].
\bibitem{magic}
  J.~M.~Drummond, J.~Henn, V.~A.~Smirnov and E.~Sokatchev,
  ``Magic identities for conformal four-point integrals,''
  JHEP {\bf 0701}, 064 (2007)
  [arXiv:hep-th/0607160].
  
\bibitem{BDS}
  Z.~Bern, L.~J.~Dixon and V.~A.~Smirnov,
  ``Iteration of planar amplitudes in maximally supersymmetric Yang-Mills
  theory at three loops and beyond,''
  Phys.\ Rev.\  D {\bf 72}, 085001 (2005)
  [arXiv:hep-th/0505205].
  
\bibitem{4pt5pt}
  J.~M.~Drummond, J.~Henn, G.~P.~Korchemsky and E.~Sokatchev,
  ``On planar gluon amplitudes/Wilson loops duality,''
  Nucl.\ Phys.\  B {\bf 795}, 52 (2008)
  [arXiv:0709.2368 [hep-th]].


\bibitem{aldaymaldacena1}
  L.~F.~Alday and J.~M.~Maldacena,
  ``Gluon scattering amplitudes at strong coupling,''
  JHEP {\bf 0706}, 064 (2007)
  [arXiv:0705.0303 [hep-th]];
\bibitem{aldaymaldacena2}
  L.~F.~Alday and J.~Maldacena,
  ``Comments on gluon scattering amplitudes via AdS/CFT,''
  JHEP {\bf 0711}, 068 (2007)
  [arXiv:0710.1060 [hep-th]].
\bibitem{LipatovRegge}
  J.~Bartels, L.~N.~Lipatov and A.~Sabio Vera,
  Phys.\ Rev.\  D {\bf 80}, 045002 (2009)
  [arXiv:0802.2065 [hep-th]].
 \bibitem{maldacenaberkovits}
  N.~Berkovits and J.~Maldacena,
  ``Fermionic T-Duality, Dual Superconformal Symmetry, and the Amplitude/Wilson
  Loop Connection,''
  JHEP {\bf 0809}, 062 (2008)
  [arXiv:0807.3196 [hep-th]].
\bibitem{aldayroiban}
  L.~F.~Alday and R.~Roiban,
  ``Scattering Amplitudes, Wilson Loops and the String/Gauge Theory
  Correspondence,''
  Phys.\ Rept.\  {\bf 468}, 153 (2008)
  [arXiv:0807.1889 [hep-th]].

\bibitem{WL1}
  J.~M.~Drummond, G.~P.~Korchemsky and E.~Sokatchev,
  ``Conformal properties of four-gluon planar amplitudes and Wilson loops,''
  Nucl.\ Phys.\  B {\bf 795}, 385 (2008)
  [arXiv:0707.0243 [hep-th]].
\bibitem{brandhuberloop}
  A.~Brandhuber, P.~Heslop and G.~Travaglini,
  Nucl.\ Phys.\  B {\bf 794}, 231 (2008)
  [arXiv:0707.1153 [hep-th]].


 \bibitem{hexagon1}
  J.~M.~Drummond, J.~Henn, G.~P.~Korchemsky and E.~Sokatchev,
  ``The hexagon Wilson loop and the BDS ansatz for the six-gluon amplitude,''
  Phys.\ Lett.\  B {\bf 662}, 456 (2008)
  [arXiv:0712.4138 [hep-th]].
\bibitem{hexagon2}
  Z.~Bern, L.~J.~Dixon, D.~A.~Kosower, R.~Roiban, M.~Spradlin, C.~Vergu and A.~Volovich,
  ``The Two-Loop Six-Gluon MHV Amplitude in Maximally Supersymmetric Yang-Mills
  Theory,''
  Phys.\ Rev.\  D {\bf 78}, 045007 (2008)
  [arXiv:0803.1465 [hep-th]].
\bibitem{hexagon3}
  J.~M.~Drummond, J.~Henn, G.~P.~Korchemsky and E.~Sokatchev,
  ``Hexagon Wilson loop = six-gluon MHV amplitude,''
  Nucl.\ Phys.\  B {\bf 815}, 142 (2009)
  [arXiv:0803.1466 [hep-th]].

\bibitem{sokatchevDCI}
  J.~M.~Drummond, J.~Henn, G.~P.~Korchemsky and E.~Sokatchev,
  ``Dual superconformal symmetry of scattering amplitudes in N=4
  super-Yang-Mills theory,''
  Nucl.\ Phys.\  B {\bf 828}, 317 (2010)
  [arXiv:0807.1095 [hep-th]].
\bibitem{brandhubernote}
  A.~Brandhuber, P.~Heslop and G.~Travaglini,
  ``A note on dual superconformal symmetry of the N=4 super Yang-Mills
  S-matrix,''
  Phys.\ Rev.\  D {\bf 78}, 125005 (2008)
  [arXiv:0807.4097 [hep-th]].
\bibitem{ahgrassmannian}
  N.~Arkani-Hamed, F.~Cachazo and C.~Cheung,
  ``The Grassmannian Origin Of Dual Superconformal Invariance,''
  JHEP {\bf 1003}, 036 (2010)
  [arXiv:0909.0483 [hep-th]].
\bibitem{masonDCI}
  L.~Mason and D.~Skinner,
  ``Dual Superconformal Invariance, Momentum Twistors and Grassmannians,''
  JHEP {\bf 0911}, 045 (2009)
  [arXiv:0909.0250 [hep-th]].

\bibitem{Hodges}
  A.~Hodges,
  ``Eliminating spurious poles from gauge-theoretic amplitudes,''
  arXiv:0905.1473 [hep-th].

\bibitem{sokatchevsDCI}
  G.~P.~Korchemsky and E.~Sokatchev,
  ``Symmetries and analytic properties of scattering amplitudes in N=4 SYM
  theory,''
  Nucl.\ Phys.\  B {\bf 832}, 1 (2010)
  [arXiv:0906.1737 [hep-th]].




\bibitem{masonskinner}
  L.~Mason and D.~Skinner,
  ``The Complete Planar S-matrix of N=4 SYM as a Wilson Loop in Twistor
  Space,''
  arXiv:1009.2225 [hep-th].


\bibitem{EKS2}
  B.~Eden, G.~P.~Korchemsky and E.~Sokatchev,
  ``More on the duality correlators/amplitudes,''
  arXiv:1009.2488 [hep-th].

\bibitem{EKS1}
  B.~Eden, G.~P.~Korchemsky and E.~Sokatchev,
  ``From correlation functions to scattering amplitudes,''
  arXiv:1007.3246 [hep-th].

\bibitem{abcct}
  N.~Arkani-Hamed, J.~L.~Bourjaily, F.~Cachazo, S.~Caron-Huot and J.~Trnka,
  ``The All-Loop Integrand For Scattering Amplitudes in Planar N=4 SYM,''
  arXiv:1008.2958 [hep-th].

\bibitem{loopandtrees}
  S.~Caron-Huot,
  ``Loops and trees,''
  arXiv:1007.3224 [hep-ph].


\bibitem{nair}
  V.~P.~Nair,
  Phys.\ Lett.\  B {\bf 214}, 215 (1988).

\bibitem{MHVt}
  M.~Bullimore, L.~Mason and D.~Skinner,
  ``MHV Diagrams in Momentum Twistor Space,''
  arXiv:1009.1854 [hep-th].

\bibitem{chalmerssiegel}
  G.~Chalmers and W.~Siegel,
  ``The self-dual sector of {QCD} amplitudes,''
  Phys.\ Rev.\  D {\bf 54}, 7628 (1996)
  [arXiv:hep-th/9606061].
  
\bibitem{elvang}
  H.~Elvang, D.~Z.~Freedman and M.~Kiermaier,
  ``Dual conformal symmetry of 1-loop NMHV amplitudes in N=4 SYM theory,''
  JHEP {\bf 1003}, 075 (2010)
  [arXiv:0905.4379 [hep-th]].

\bibitem{inprep}
 N.~Arkani-Hamed, J.~Bourjaily, F.~Cachazo, S.~Caron-Huot and J.~Trnka, in preparation.

\bibitem{sDCI2}
  G.~P.~Korchemsky and E.~Sokatchev,
  ``Superconformal invariants for scattering amplitudes in N=4 SYM theory,''
  Nucl.\ Phys.\  B {\bf 839}, 377 (2010)
  [arXiv:1002.4625 [hep-th]];
  J.~M.~Drummond and L.~Ferro,
  ``The Yangian origin of the Grassmannian integral,''
  arXiv:1002.4622 [hep-th].

\bibitem{Aldayetal}
  L.~F.~Alday, B.~Eden, G.~P.~Korchemsky, J.~Maldacena and E.~Sokatchev,
  ``From correlation functions to Wilson loops,''
  arXiv:1007.3243 [hep-th].

\bibitem{BCF}
  R.~Britto, F.~Cachazo and B.~Feng,
  ``New Recursion Relations for Tree Amplitudes of Gluons,''
  Nucl.\ Phys.\  B {\bf 715}, 499 (2005)
  [arXiv:hep-th/0412308].
\bibitem{BCFW}
  R.~Britto, F.~Cachazo, B.~Feng and E.~Witten,
  ``Direct Proof Of Tree-Level Recursion Relation In Yang-Mills Theory,''
  Phys.\ Rev.\ Lett.\  {\bf 94}, 181602 (2005)
  [arXiv:hep-th/0501052].

\bibitem{drummondhenn}
  J.~M.~Drummond and J.~M.~Henn,
  JHEP {\bf 0904}, 018 (2009)
  [arXiv:0808.2475 [hep-th]].



\bibitem{intriligator}    
  K.~A.~Intriligator,
  ``Bonus symmetries of N = 4 super-Yang-Mills correlation functions via  AdS
  duality,''
  Nucl.\ Phys.\  B {\bf 551}, 575 (1999)
  [arXiv:hep-th/9811047].
 \bibitem{emerynonrenorm}
  B.~Eden, A.~C.~Petkou, C.~Schubert and E.~Sokatchev,
  ``Partial non-renormalisation of the stress-tensor four-point function in  N
  = 4 SYM and AdS/CFT,''
  Nucl.\ Phys.\  B {\bf 607}, 191 (2001)
  [arXiv:hep-th/0009106].
\bibitem{emery2loops}
  B.~Eden, C.~Schubert and E.~Sokatchev,
  ``Three-loop four-point correlator in N = 4 SYM,''
  Phys.\ Lett.\  B {\bf 482}, 309 (2000)
  [arXiv:hep-th/0003096].
\bibitem{contactterms}    
  A.~Petkou and K.~Skenderis,
  ``A non-renormalization theorem for conformal anomalies,''
  Nucl.\ Phys.\  B {\bf 561}, 100 (1999)
  [arXiv:hep-th/9906030].
\bibitem{howe}
  P.~S.~Howe, C.~Schubert, E.~Sokatchev and P.~C.~West,
  ``Explicit construction of nilpotent covariants in N = 4 SYM,''
  Nucl.\ Phys.\  B {\bf 571}, 71 (2000)
  [arXiv:hep-th/9910011].



  
\bibitem{weinberg}
  S.~Weinberg,
  Phys.\ Rev.\  D {\bf 82}, 045031 (2010)
  [arXiv:1006.3480 [hep-th]].

\bibitem{volovich}
  A.~B.~Goncharov, M.~Spradlin, C.~Vergu and A.~Volovich,
  ``Classical Polylogarithms for Amplitudes and Wilson Loops,''
  arXiv:1006.5703 [hep-th].
 

\bibitem{aldayhenn}
  L.~F.~Alday, J.~M.~Henn, J.~Plefka and T.~Schuster,
  ``Scattering into the fifth dimension of N=4 super Yang-Mills,''
  JHEP {\bf 1001}, 077 (2010)
  [arXiv:0908.0684 [hep-th]].
\bibitem{henn4loop}
  J.~M.~Henn, S.~G.~Naculich, H.~J.~Schnitzer and M.~Spradlin,
  ``More loops and legs in Higgs-regulated N=4 SYM amplitudes,''
  JHEP {\bf 1008}, 002 (2010)
  [arXiv:1004.5381 [hep-th]].
\bibitem{bern2loop}
  Z.~Bern, J.~S.~Rozowsky and B.~Yan,
  ``Two-loop four-gluon amplitudes in N = 4 super-Yang-Mills,''
  Phys.\ Lett.\  B {\bf 401}, 273 (1997)
  [arXiv:hep-ph/9702424].


\end{thebibliography}
\end{document}